\documentclass[a4paper,11pt]{article}

\usepackage[width=15cm,height=23cm]{geometry} 
\usepackage[usenames]{color}
\usepackage{amsfonts,amssymb}
\usepackage{theorem}
\usepackage[dvips]{graphicx}

\long\def\symbolfootnote[#1]#2{\begingroup%
\def\thefootnote{\fnsymbol{footnote}}\footnote[#1]{#2}\endgroup}

\begin{document}
\centerline{}
\vskip 2.4 truecm
\normalsize
\begin{center}{\Large\bf
Metamorphosis versus Decoupling in Nonabelian Gauge \\Theories  at Very High Energies
}
\end{center}

\par

\large
\rm
\vskip 0.7 truecm
\centerline{ 
Ruggero~Ferrari$^{a,b}$
\footnote{e-mail: {\tt ruggferr@mit.edu}}}

\small
\medskip
\begin{center}
$^a$
Center for Theoretical Physics\\
Laboratory for Nuclear Science\\
and Department of Physics\\
Massachusetts Institute of Technology\\
Cambridge, Massachusetts 02139\\
and\\
$^b$
Dip. di Fisica, Universit\`a degli Studi di Milano\\
and INFN, Sez. di Milano\\
via Celoria 16, I-20133 Milano, Italy\\
(MIT-CTP 4275, IFUM-977-FT, June, 2011 )
\end{center}

\normalsize


\begin{quotation}

\rm
 {\Large Abstract:}
 In the present paper we study the limit of zero mass in nonabelian
 gauge theories both with Higgs mechanism and in the nonlinear realization
 of the gauge group (St\"uckelberg mass).  We argue that in the first
 case the longitudinal modes undergo a metamorphosis process to the
 Goldstone scalar modes, while in the second we guess a decoupling
 process associated to a phase transformation.
 \par\noindent
 The two scenarios yield strikingly different  behaviors at high energy,
 mainly ascribed to the presence of a massless Higgs doublet
 among the physical modes in the case of Higgs mechanism (i.e. not only
 the Higgs boson). 
 \par\noindent
 The aim of this work is to show that the problem of unitarity at high energy
 in nonabelian gauge theory with no Higgs boson can open new perspectives
 in quantum field theory.

\end{quotation}

\newpage
\section{Introduction}
The fate of the longitudinal mode of a vector boson at high energy
is entangled with the problem of unitarity. On the basis of some
well known  papers in the late 70's and early 80's 
\cite{Cornwall:1974km}-\cite{Chanowitz:1985hj}
people have acquired the conviction that the Higgs boson is necessary
in order to ensure physical unitarity in nonabelian gauge theories. 
The heart of the argument is 
based on the behavior of the elastic scattering amplitude of the
longitudinal modes $W_L~ W_L$ at high energy 
\cite{Cornwall:1974km}-
\cite{Denner:1997kq}.

Thus the construction of Higgsless Elecroweak Models faces
tremendously difficult theoretical problems dealing with basic
principles as renormalization, unitarity, foundation of bound state
quantum field theory, predictivity of the model (finite number
of free parameters), etc. Ref. \cite{referee} updates a recent proposal
for a Higgless scenario and provides a nice overview of most models.
\par
The problem comes from the fact that the longitudinal polarization
of a vector meson is a physical mode for any finite
value of the mass ($M$). On the other side, for zero mass a vector meson
has only two (transverse) polarizations. Thus either the mode decouples
from the physical states
in the massless limit (like in massive QED) or we face a singular behavior
at zero vector boson mass.
\par
In nonabelian gauge theories (we deal with $SU(2)$, with or without the $U(1)$
factor) 
the longitudinal polarization
does not decouple in physical S-matrix elements at
zero mass.
A conundrum that shows up in really practical items as in the proof of
physical unitarity and in phenomenology\cite{Slavnov:1970tk,Curci:1976kh}.
The number of physical modes changes in the limit of zero
mass, thus the cancellation of the unphysical modes in the
proof of S-matrix unitarity
must proceed with different patterns in the two regimes. On the other side,
in phenomenology one must introduce a cut-off to mark the 
events region, where the longitudinal polarization of the vector meson
can be established within the errors (at very high momentum one
cannot distinguish between the longitudinal polarization and the spin zero
wave function). Therefore the distinction between longitudinal mode for the spin one
and a spin zero mode, in the limit of zero mass, has no operative meaning
and, as said, conflicts with unitarity, due to the non-decoupling.
\par
A na{\"\i}ve analysis, based on BRST transformations, indicates that
in the massless limit 
the {\sl unphysical} (at $M\not =0$) components  of the Higgs field 
eventually describe  {\sl physical} massless scalars for $M =0$.
Thus we suggest that the longitudinal polarization mode,
in default of the decoupling, undergoes a metamorphosis to the
mode of a massless scalar.  Physical unitarity of the S-matrix is preserved
with this assignment of the fields to physical and unphysical
modes. The Equivalence Theorem (CLTCG)
\cite{Cornwall:1974km}, \cite{Chanowitz:1985hj}, \cite{Gounaris:1986cr},
\cite{Yao:1988aj}-\cite{Denner:1996gb}
supports  this setting.
\footnote{
After the work of 
J.~M.~Cornwall, D.~N.~Levin, G.~Tiktopoulos,
  M.~S.~Chanowitz and M.~K.~Gaillard the relation
  between the S-matrix elements for longitudinal modes
  of the gauge fields
  and those of the Goldstone bosons has developed to a
  somewhat more complex result, than the one implied 
  by the theorem on the point transformations of fields in scattering theory
  \cite{et}.
  The present work adds more consequences to the 
  discovery of the above mentioned physicists. Thus
  we choose to denote the theorem by their names. 
}
\par\noindent
Of course the massless theory is plagued by infrared divergences;
but we are going to ignore this difficulty, hoping
that the two problems are not dangerously entangled. If instead infrared
divergences are an insurmountable obstacle, as a last resource one can
reverse the view point and consider the mass as the infrared regulator
of an otherwise ill-defined field theory.
\par
This possibility of a metamorphosis of states is suggested for 
nonabelian gauge theories
where the mass $M$ is generated by the Higgs mechanism. 
The reason being that the limit of symmetry restoration
(zero vacuum expectation value (v.e.v) of the Higgs boson field) seems
doable on the effective action in a loop-wise perturbative expansion.
\par
After the metamorphosis, the theory consists of a massless gauge field
and a complex doublets of scalar fields (the fields used to induce the Higgs
mechanism for $M\not =0$). According to the standard analysis the theory
is not asymptotically free \cite{Gross:1973ju}-\cite{Politzer:1973fx},
due to the presence of scalars. 
\par
Thus one gets
a consistent setting for studying the  physics of  the intermediate
vector mesons at energies $E>>M_W,M_Z$.  If needed, one
can use the $M\not =0$ theory as the infrared regulated theory.
\par
In the present paper we address the same question 
in the case where $M$ enters via a  St\"uckelberg
term.
In this case the local gauge group is realized  \underline{nonlinearly} 
and hence there is no need of a Higgs boson in the perturbative spectrum.
\par
Both theories obey the same set of equations used in the present work: Slavnov-Taylor Identity
(STI), gauge fixing  equation and anti-ghost equation
\footnote{
The Local Functional Equation \cite{Ferrari:2005ii}, employed in the subtraction
strategy of the ultraviolet divergences in nonlinear theories, is not used
here.}.
Moreover the CLTCG theorem takes the same form. However
they are strikingly different in the zero mass limit. In the linear case the limit
of v.e.v  to zero in the Feynman amplitudes seems to be manageable.
While in the nonlinear case the situation is fuzzier. In recent works
\cite{Ferrari:2005ii}-\cite{Ferrari:2009uj} we proposed a divergences subtraction
scheme for the nonlinear sigma model, for the massive Yang-Mills
theory and for the Electroweak Model $SU(2)\otimes U(1)$.
Locality and perturbative unitarity are obeyed. However the perturbative 
solution has a bad $M^{-1}$ behavior for $M=0$, essentially due to the 
nonlinear sigma model couplings.
\par\noindent
Thus the scenario of a metamorphosis of the longitudinal modes
for $M=0$ cannot be envisaged in the case of nonlinear realization of the
gauge group.
\par
From some considerations, based on the matching of the
number of degrees of freedom and on the strong coupling limit of
the lattice-regulated theory, we guess a $M\to 0$ behavior where
both the longitudinal polarizations and the Goldstone bosons decouple. 
In order  to support this scenario,
we envisage the existence of two or more phases in the parameter
space separated by some discontinuity. In particular we assume that
the loop expansion cannot be continued to $M=0$.
This would mark the difference with the linear case, where the Goldstone
bosons survive as physical modes. For instance the nonlinear theory would
be an asymptotically free theory in the limit.
\par\noindent
The conjecture on the limit $M=0$ for the nonlinear case could be
studied by lattice simulations. In particular one should make a survey
of the phase diagram in the parameter space $(g^{-2},M^2)$ and look for possible
singularities  responsible for the bad behavior of the loop expansion for low
mass
\footnote{
Recent lattice simulations \cite{Ferrari:2011ef} support this scenario
}.
\par\noindent
A further comparison of the two scenarios could come from high energy processes.
However a quick analysis shows that this is pretty hard to achieve, as a simple
example will show.
\par\noindent
We work in the 't Hooft  gauge.
\section{The Classical Actions}
\label{sec:action}
In this section we fix some notations. Matter fields are omitted
in most part of the paper.
\par\noindent
The work  focuses on the gauge and scalar sectors of the following Yang-Mills
classical actions written for the $SU(2)$ gauge group.
We consider both cases of linear (Higgs) and nonlinear
(St\"uckelberg) representation of the gauge group.
\subsection{Yang Mills with Higgs Mechanism }
\label{sec:YM}
We consider a $SU(2)$ Yang-Mills theory where the mass
is generated through the Higgs mechanism
\begin{eqnarray}
S_{\rm  H}  = \frac{\Lambda^{(D-4)}}{g^2} \int d^Dx \, 
\biggl( -\frac{1}{4} G_{a\mu\nu} G_a^{\mu\nu}
 +\Big [(\partial_\mu -i A_\mu) \Phi\Big]^\dagger
\Big[(\partial^\mu -i A^\mu) \Phi\Big] 
-\frac{\lambda}{4g^2}(\Phi^\dagger\Phi -2v^2g^2 )^2
\biggr) .
\label{YM.1}
\end{eqnarray}
$\Lambda$ is a mass scale for the analytic continuation in $D$ dimensions. We use the
short notation
\begin{eqnarray}
\Lambda_g\equiv \frac{\Lambda^{(D-4)}}{g^2} .
\label{YM.1.1}
\end{eqnarray}
We use the matrix notation
\begin{eqnarray}&&
A\mu= \frac{\tau_a}{2}A_{a\mu}
\nonumber \\&&
G_{\mu\nu}[A] = G_{a\mu\nu} \frac{\tau_a}{2} =
\partial_\mu A_\nu - \partial_\nu A_\mu -i[A_\mu,A_\nu]
\label{YM.2}
\end{eqnarray}
and $\Phi$ is parametrized by
\begin{eqnarray}
\Phi=
\frac{1}{\sqrt 2}
\left(
\begin{array}{l}
i\phi_1+\phi_2\\
\phi_0 -i\phi_3
\end{array}
\right).
\label{YM.3}
\end{eqnarray}
The action (\ref{YM.1}) is invariant 
under  local $SU(2)_L$ left transformations 
\begin{eqnarray}
&& A'_\mu = U_L A_\mu U_L^\dagger + i U_L \partial_\mu U_L^\dagger 
\nonumber \\ &&
\Phi' = U_L\Phi
\label{YM.3.1}
\end{eqnarray}
and under global $SU(2)_R$ right transformations 
\begin{eqnarray}
&& A'_\mu = A_\mu 
\nonumber \\ &&
\Omega' = \Omega \, U_R,
\label{YM.3.2}
\end{eqnarray}
where
\begin{eqnarray}
&& 
\Omega_{\alpha \beta} \simeq \sqrt 2 \Phi_\alpha \tilde\Phi_\beta
\nonumber \\ &&
\tilde\Phi = i \tau_2 \Phi^*.
\label{YM.3.3}
\end{eqnarray}
Notice that in general $\Omega \not\in SU(2)$.
\par
The spontaneous breakdown of the global $SU(2)_L\otimes SU(2)_R$ symmetry proceeds
via the nonzero vacuum expectation value
\begin{eqnarray}
\langle 0|\phi_0|0\rangle=\langle 0|(h+2vg)|0\rangle=2vg
\label{YM.4}
\end{eqnarray}
and the global $SU(2)$ invariance is left over on the vector indexes. 
\par\noindent
The spontaneous breakdown induces a mass for the vector bosons ($M\equiv gv$), for the Higgs 
boson ($M^2_H \equiv \lambda v^2 $) and a mixing
$A_\mu - \phi$
\begin{eqnarray}&&
S_{\rm H\, Bilinear}  = \frac{\Lambda^{(D-4)}}{g^2} \int d^Dx \, 
\biggl( -\frac{1}{2}(\partial_\mu A_{a\nu} \partial^\mu A_a^\nu
-\partial_\mu A_{a\nu} \partial^\nu A_a^\mu)
\nonumber \\&&
 + \frac{M^2}{2}A_{a\nu} A_a^\nu 
+ \frac{1}{2}\partial_\mu h\partial^\mu h
 + \frac{1}{2}\partial_\mu\phi_a\partial^\mu\phi_a
- M^2A_{a\mu}\partial^\mu\phi_a - \frac{1}{2}M^2_{H} h^2\biggr).
\label{YM.5}
\end{eqnarray}
We use the 't Hooft gauge in order to remove the mixing. In doing this we get a mass
for the Goldstone bosons $\vec\phi$
\begin{eqnarray}
S_{\rm  H\,gf}  = \frac{\Lambda^{(D-4)}}{g^2} \int d^Dx \, 
\biggl( \frac{b^2}{2\xi }+\frac{ M}{\xi }b_a\phi_a +
b_a  \partial_\mu A_a^\mu \biggr). 
\label{YM.6}
\end{eqnarray}
The Goldstone mass at the tree level is then $M_{G}^2\equiv  M^2 \xi^{-1 }$.
In Appendix \ref{app:app1} we give the complete effective action at zero
loop with the necessary Faddeev-Popov ghosts.
\par
The perturbation expansion is in the number of loops and the amplitudes are made
finite by using na{\"\i}ve dimensional renormalization. Finite renormalization
is not a relevant item for the content of the paper. 
\subsection{Yang-Mills with St\"uckelberg Mass}
The nonlinear sigma model field $\Omega$ is an element of the $SU(2)$ group, 
which is parametrized in terms of the
coordinate fields $\phi_a$ as follows (compare with eq. (\ref{YM.3.3}))
\footnote{In the nonlinear case we use dimensionless fields $\phi_a$.}
\begin{eqnarray}
&& \Omega =  \phi_0 + i \tau_a \phi_a  \, , ~~~ \Omega^\dagger \Omega = 1 \,  , ~~~ {\rm det} \, \Omega = 1 \, ,
\nonumber \\
&& \phi_0^2 + \phi_a^2 = 1  \, .
\label{s2.2}
\end{eqnarray}
The $SU(2)$ flat connection is 
\begin{eqnarray}
&& F_\mu = i \Omega \partial_\mu \Omega^\dagger 
= F_{a\mu} \frac{\tau_a}{2} \, , \nonumber \\
&& F_{a\mu} = 2 (\phi_0 \partial_\mu \phi_a -
\partial_\mu \phi_0 \phi_a + \epsilon_{abc} \partial_\mu \phi_b \phi_c ) \, .
\label{s2.1}
\end{eqnarray}
The field strength of $F_\mu$ vanishes 
\begin{eqnarray}
G_{\mu\nu}[F] = 0 \, .
\label{flc.1}
\end{eqnarray}
Under a local $SU(2)$ left transformation 
$U_L = \exp \Big ( i \alpha^L_a \frac{\tau_a}{2} \Big )$ one gets
\begin{eqnarray}
&& \Omega' =  U_L \Omega \, , \nonumber \\
&& F'_\mu = U_L F_\mu U_L^\dagger + i U_L \partial_\mu U_L^\dagger \, , \nonumber \\
&& A'_\mu = U_L A_\mu U_L^\dagger + i U_L \partial_\mu U_L^\dagger \, .
\label{s2.4}
\end{eqnarray}
The constraint in eq.(\ref{s2.2})
implies that the gauge symmetry is nonlinearly realized on the 
fields $\phi_a$, whose
infinitesimal transformations are
\begin{eqnarray}
&& \delta \phi_a = \frac{1}{2} \phi_0 \alpha^L_a + \frac{1}{2} \epsilon_{abc} \phi_b \alpha^L_c \, , ~~~~
\phi_0 = \sqrt{1  - \phi_a^2} \, , \nonumber \\
&& \delta \phi_0 = -\frac{1}{2} \alpha^L_a \phi_a \, .
\label{s2.4.1}
\end{eqnarray}
Under local $SU(2)_L$ symmetry the combination $A_\mu - F_\mu$ transforms in the adjoint representation of $SU(2)$.
Hence one can construct out of $A_\mu - F_\mu$ and $\Omega$ invariants under 
$SU(2)_L$ local transformations.
The Yang-Mills
action in the presence of a St\"uckelberg mass term \cite{Bettinelli:2007tq} 
and in the 't Hooft gauge is
\begin{eqnarray}&&
S_{\rm S}  = \frac{\Lambda^{(D-4)}}{g^2} \int d^Dx \, \Big ( - \frac{1}{4} 
                              G_{a\mu\nu}[A] G^{\mu\nu}_a[A] + 
                    \frac{M^2}{2} (A_{a\mu} - F_{a\mu})^2  \Big ) \,
\nonumber\\&&
S_{\rm  S\,gf}  = \frac{\Lambda^{(D-4)}}{g^2} \int d^Dx \, 
\biggl( \frac{b^2}{2\xi }+2\frac{ M^2}{\xi }b_a\phi_a +
b_a  \partial_\mu A_a^\mu \biggr).
\label{stck.1}
\end{eqnarray}
$S_S$ is invariant under local $SU(2)_L$ symmetry  and also global $SU(2)_R$
symmetry. The choice of independent fields made in eq. (\ref{s2.4.1}) fixed the
direction of the spontaneous breakdown of the symmetry.
The bilinear part of the action $S_S$ is essentially the same as in the
Higgs mechanism (\ref{YM.5}) apart from the absence of the Higgs boson terms.
\section{Properties of the Two-point Functions (Higgs)}
\label{sec:ptpf}
In Appendix \ref{sec:tpf} we derive the two-point connected functions
of the unphysical bosonic sector. The solutions are given in terms of the  1PI  
two-point functions
\footnote{
We drop internal indexes whenever there is no ambiguity. Moreover
we use the notation
\begin{equation}
\Gamma_{ A^\mu A^\nu}= \Gamma_T (p^2)(g_{\mu\nu}-\frac{p_\mu p_\nu}{p^2})+
\Gamma_L (p^2) \frac{p_\mu p_\nu}{p^2}.
\label{ptpf.1}
\end{equation}
}
\begin{equation}
\Gamma_{\phi\phi}, \qquad ip^\nu\Gamma_{\phi A^\nu}, \qquad \Gamma_L
\label{ptpf.2}
\end{equation}
which are related by the eq. (\ref{tpf.9.2})
\begin{equation}
(p^\nu\Gamma_{A^\nu\phi})^2 +p^2\Gamma_L\Gamma_{\phi\phi}=0.
\label{ptpf.3}
\end{equation}
The  connected two-point functions involving the Lagrangian multiplier
are (see eq. (\ref{basic.8}))
\begin{eqnarray}&&
 W_{A^\mu b}=-\frac{i}{\Lambda_g}\frac{p^\mu}{
p^2-i  \frac{M}{\xi}\frac{p^\nu\Gamma_{\phi A^\nu}}{\Gamma_{\phi\phi}}
}
\nonumber\\&&
 W_{\phi b}
=\frac{i}{\Lambda_g}\frac{p^\nu\Gamma_{\phi A^\nu}}{\Gamma_{\phi\phi}}
\frac{1}{p^2-\frac{M}{\xi}\frac{ip^\nu\Gamma_{\phi A^\nu}}{\Gamma_{\phi\phi}}
}.
\label{ptpf.4}
\end{eqnarray}
The two-point functions have the pole in the same position, i.e. the solution 
of
\begin{eqnarray}
p^2=\frac{M}{\xi}\frac{ip^\nu\Gamma_{\phi A^\nu}}{\Gamma_{\phi\phi}}.
\label{basic.99}
\end{eqnarray}
\par
The connected two-point functions for unphysical modes involving $ \phi$ and $ A^\mu$ are 
given in eqs. (\ref{tpf.22}), (\ref{tpf.23}) and (\ref{tpf.24})
\begin{eqnarray}
W_{\phi \phi}
 =-
\frac{p^2}{ \Gamma_{\phi \phi}}
\biggl(p^2-
\frac{1}{\Lambda_g\xi}\Gamma_L 
\biggr)
\frac{1}{\biggl(
p^2 -  \frac{M}{\xi}
\frac{ip^\nu\Gamma_{\phi A^\nu}}{\Gamma_{\phi\phi}}\biggr)^2
},
\label{ptpf.22}
\end{eqnarray}
\begin{eqnarray}
  W_{A^\mu \phi}
=i \frac{1}{\xi\Gamma_{\phi\phi}}\biggl(\frac{i}{\Lambda_g}p^\nu\Gamma_{ A^\nu\phi}
+M p^2
\biggr)
\frac{p^\mu
}{\biggl(
p^2 -  \frac{M}{\xi} 
\frac{ip^\nu\Gamma_{ \phi A^\nu}}{\Gamma_{\phi\phi}}\biggr)^2
}
\label{ptpf.23}
\end{eqnarray}
and
\begin{eqnarray}
W_L 
=\frac{p^2}{\xi\Gamma_{\phi\phi}}
\frac{\frac{\Gamma_{\phi\phi}}{\Lambda_g} -\frac{M^2}{\xi}}{\biggl(
p^2 -  \frac{M}{\xi} \frac{ip^\nu\Gamma_{\phi A^\nu}}
{\Gamma_{\phi\phi}}\biggr)^2
}.
\label{ptpf.24}
\end{eqnarray}
By direct computation one can derive the identity
\begin{eqnarray}
\Gamma_L = \frac{p^2 W_{ b\phi}^2}{  \frac{1}{\xi}    W_{ b\phi}^2-\frac{1}{\Lambda_g}W_{\phi\phi}   } \, .
\label{ptpf.24.1}
\end{eqnarray}
It is interesting to see the properties of the numerators in eqs. (\ref{ptpf.22}), 
(\ref{ptpf.23})  and (\ref{ptpf.24}). They form a matrix 
(variables: $\phi,M^{-1}\partial^\mu A_\mu$)
\begin{eqnarray}
{\cal G}\equiv
\frac{p^2}{ \xi\Gamma_{\phi \phi}}
\left(
\begin{array}{ll}-\xi p^2 +\frac{\Gamma_L}{\Lambda_g} &M^{-1}(\frac{i}{\Lambda_g}p^\nu\Gamma_{ A^\nu\phi}
+M p^2)\\
M^{-1}(\frac{i}{\Lambda_g}p^\nu\Gamma_{ A^\nu\phi}
+M p^2) & M^{-2}(p^2 \frac{\Gamma_{\phi\phi}}{\Lambda_g} -p^2 \frac{M^2}{\xi})
\end{array}
\right)
\label{ptpf.26}
\end{eqnarray}
whose determinant is
\begin{eqnarray}
\det ({\cal G})
=-\frac{1}{\Lambda_g}\frac{p^4}{M^2 \xi\Gamma_{\phi \phi}}\biggl( p^2 
- \frac{M}{\xi}\frac{ip^\nu\Gamma_{\phi A^\nu}}{\Gamma_{\phi \phi}}\biggr)^2.
\label{ptpf.27}
\end{eqnarray}
Thus the numerator-matrix  has one vanishing eigenvalue  on the double-poles.
In fact the field given by the linear combination \cite{Becchi:1974md}
\begin{eqnarray}
X_1 = \frac{\phi}{\xi} + \frac{\partial^\mu A_\mu}{M}
\label{ptpf.28}
\end{eqnarray}
is shown  (Appendix \ref{app:conv} eq.(\ref{tpf.49}))  to be the eigenvector of the vanishing eigenvalue of the
matrix in eq. (\ref{ptpf.26}).
%
\section{The Na{\"\i}ve $M=0$ Limit (Higgs)}
\label{sec:limh}
The paper is focused on extreme processes where one can
neglect the mass parameters. In the Higgs case this is equivalent
to the limit $v=0$, i.e. the limit of unbroken symmetry.
\par
We do not consider skew limits as $M=0$ and $M_H \not =0$, which,
although interesting in phenomenology \cite{Denner:1996gb}, requires a series resummation
as $v\to 0$.
\par 
In the limit $v=0$  the symmetry $\Phi \to -\Phi$ is unbroken; as a
consequence one has
\begin{eqnarray}
\Gamma_{\phi A^\nu} =0
\label{ptpf.29}
\end{eqnarray}
and therefore  
\begin{eqnarray}&&
\Gamma_{L}=0
\nonumber\\&&
 W_{A^\mu b}=-i\frac{p^\mu}{\Lambda_g
p^2
}
\nonumber\\&&
 W_{\phi b}
=0.
\label{ptpf.31.0}
\end{eqnarray}
Similarly the limit in the eqs. (\ref{ptpf.23}), (\ref{ptpf.22}) and (\ref{ptpf.24})
yields
\begin{eqnarray}&&
 W_{A^\mu \phi}
=0
\nonumber\\&&
W_{\phi \phi}
 =-
\frac{1}{ \Gamma_{\phi \phi}}
\nonumber\\&&
W_L 
=\frac{1}{\Lambda_g\xi}
\frac{1}{p^2 }.
\label{ptpf.31.3}
\end{eqnarray}
The vanishing of the two-point functions $ W_{A^\mu \phi}$ and
$ W_{b \phi}$ shows that in the limit $\phi$ describes modes orthogonal
to the unphysical modes. This fact is a preliminary condition 
for the realization of the metamorphosis of the vector meson longitudinal 
polarizations into the Goldstone bosons, as described in the next Sections.
Moreover this suggests that the limit $v=0$ can be performed order by order 
in perturbation theory. This limit is possible on the amplitudes 
for generic external momenta, while
for most S-matrix elements the limit cannot be performed due
 to infrared divergences.
The amplitudes are those of massless nonabelian field coupled 
to a massless fields
$\Phi$ belonging to the spinorial representation of the 
$SU(2)$ group of local left transformations. 
\section{The Longitudinal Polarization and its Fate for $M\to 0$}
\label{sec:long}
The longitudinal polarization
\footnote{In this Section, and in the sequel, ${\tilde M}$ and ${\tilde M}_G$
are fixed by the poles of the transverse and longitudinal tensors
of the vector mesons after radiative corrections.}
\begin{eqnarray}
\epsilon_L = \frac{1}{{\tilde M} }\Big(|\vec p|,\frac{\vec p}{|\vec p|}E\Big), 
\qquad E=\sqrt{{\tilde M}\,^2+\vec p\,^2}
\label{long.1}
\end{eqnarray}
can be written ($E_G\equiv \scriptstyle{\sqrt{\vec p\,^2+{\tilde M}_G\,\,^2}}$)
\begin{eqnarray}
\epsilon_L = \frac{1}{{\tilde M}}\Big(E_G, \vec p\Big)
+  \Big ( - \frac{\tilde M_G\,^2}{\tilde M}\frac{1}{(p+E_G)},  \frac{\tilde M}{p(p+E)}\vec p\Big).
\label{long.2}
\end{eqnarray}
Thus for large value of the energy ($E>>{\tilde M}$) we have
\begin{eqnarray}
\epsilon_L = \frac{1}{{\tilde M}}\Big(E_G, \vec p\Big)
+ {\cal O}(\frac{{\tilde M}}{E}).
\label{basic.11.3}
\end{eqnarray}
Equation (\ref{basic.11.3}) has attracted the attention of many physicists.
We briefly add our comments.
\subsection{The Need of a Cut-off}
Equation (\ref{basic.11.3}) shows that at very high energy one cannot
experimentally distinguish the longitudinal mode of a vector field from
a spin zero state described by a field like $\partial_\mu \phi$. 
Therefore a cut-off energy
$E_C$ should be quoted in the experimental data saying for what energy $E<<E_C$
it is possible to distinguish the two states. For $E\geq E_C$ a statement 
about the spin content (spin one longitudinal versus spin zero) of the mode is void.
The necessity of such cut-off energy  is very relevant for our problem. 
In fact, if one is interested in the dynamics of the model at
$M=0$, whose S-matrix elements are plagued by infrared divergences, the cut-off can
be used in order to evaluate the physically relevant observables. Thus the model at $M=0$
can be traded with a massive nonabelian gauge theory with Higgs mechanism, 
provided that the mass $M$ is small enough for the given kinematic setup.
%
\subsection{Default of Decoupling of Longitudinal Mode for Nonabelian Gauge Theories}
It might be tempting to neglect the ${\cal O}(\frac{{\tilde M}}{E})$ parts
in eq. (\ref{basic.11.3}) and to perform the replacement
\begin{eqnarray}
\epsilon_L\, \to \,\frac{1}{{\tilde M}}\Big(E_G, \vec p\Big)
\label{long.3}
\end{eqnarray}
in the Feynman amplitudes.
Were it possible without ambiguity, then the problem of the decoupling of the
longitudinal mode would be much easier.
Unfortunately this procedure is not allowed since mixed terms in quadratic 
or higher-order forms give finite contributions, which cannot be neglected without 
further scrutiny. The reason is connected to the validity of the condition
\begin{eqnarray}
E_G~{\mathfrak M}_0(E_G,\vec p)- p_i~{\mathfrak M}_i(E_G,\vec p)=0,
\label{long.4}
\end{eqnarray}
as it will be illustrated here to some extent.
For instance, let us consider the situation where $\epsilon_L^\mu$ multiplies 
some amplitude ${\mathfrak M}_\mu$
which depends on the momentum $p_\nu$. The limit $\tilde M =0$ might be performed 
by evaluating the difference
\begin{eqnarray}&&
\epsilon_L^\mu{\mathfrak M}_\mu(E,\vec p) - \frac{1}{{\tilde M}}\Big(
E_G~{\mathfrak M}_0(E_G,\vec p)- p_i~{\mathfrak M}_i(E_G,\vec p)\Big)
\label{long.5}
\end{eqnarray}
which is of order ${\cal O}(\frac{{\tilde M}}{E})$ as a standalone expression. 
Let us  expand around $E_G$, where it is much easier to use the STI. 
We get (the common dependence from $\vec p$ being suppressed)
\begin{eqnarray}&&
\epsilon_L^\mu{\mathfrak M}_\mu(E) - \frac{1}{{\tilde M}}p^\mu {\mathfrak M}_\mu(E_G)
=\frac{1}{{\tilde M}} \Bigg( [p-E_G]{\mathfrak M}_0(E_G)
- p_i (\frac{E}{p}-1){\mathfrak M}_i(E_G)
\nonumber\\&&
+(E-E_G)\Big\{ p\frac{\partial}{\partial E_G}{\mathfrak M}_0(E_G) - p_i \frac{E}{p}
\frac{\partial}{\partial E_G}{\mathfrak M}_i(E_G)
\Big \}
\Bigg )+ {\cal O}({\tilde M}\, ^3).
\label{long.5.1}
\end{eqnarray}
Now we use the relations
\begin{eqnarray}&&
\frac{E_G}{p} = 1 + {\cal O}({\tilde M}\, ^2)
\nonumber \\&&
\frac{E}{p} = 1 + {\cal O}({\tilde M}\, ^2)
\nonumber \\&&
E_G\frac{\partial}{\partial E_G}{\mathfrak M}_0(E_G)= \frac{\partial}{\partial E_G}E_G{\mathfrak M}_0(E_G)
- {\mathfrak M}_0(E_G)
\label{long.5.1.1}
\end{eqnarray}
and we get
\begin{eqnarray}&&
\epsilon_L^\mu{\mathfrak M}_\mu(E) - \frac{1}{{\tilde M}}p^\mu {\mathfrak M}_\mu(E_G)
=\frac{1}{{\tilde M}} \Big( [p-E]\,{\mathfrak M}_0(E_G)
- p_i (\frac{E}{p}-1){\mathfrak M}_i(E_G)
\nonumber\\&&
+(E-E_G)\Big\{ \frac{\partial}{\partial E_G}E_G{\mathfrak M}_0(E_G) - 
\frac{\partial}{\partial E_G}p_i {\mathfrak M}_i(E_G)
\Big \}
\Big )+{\cal O}({\tilde M}\, ^3). 
\nonumber\\&&
= - \frac{{\tilde M}}{2p}\Big( p \,{\mathfrak M}_0(E_G)
+ p_i {\mathfrak M}_i(E_G)\Big )
+ \frac{(E-E_G)}{{\tilde M}} \frac{\partial}{\partial E_G} [E_G{\mathfrak M}_0(E_G) -p_i {\mathfrak M}_i(E_G)] 
\nonumber\\&&
+ {\cal O}({\tilde M}\, ^3).
\label{long.5.1.2}
\end{eqnarray}
Thus, if eq. (\ref{long.4}) is valid,  the derivative term can be neglected.
\par
However the expression in eq. (\ref{long.5.1.2}) may enter into some product with terms
of order $\tilde M^{-1}$ (as for instance $\epsilon_L^\mu$ ) thus producing a non vanishing result.
Typically this happens by evaluating the sum over final states as for instance
\begin{eqnarray}
-\frac{d^3p}{2E}\,\,\, {\mathfrak M}^*_\mu \,\,\,\epsilon_L^{*\mu} \epsilon_L^{\nu} \,\,\,{\mathfrak M}_\nu,
\label{long.5.3}
\end{eqnarray}
where the ${\cal O}(\tilde M)$ term in eq. (\ref{long.5}) gives a finite contribution
when multiplied by $\tilde M^{-1}$  in $\epsilon_L^{*\mu} $. Finally the problem
consists in evaluating  
\begin{eqnarray}
\frac{1}{\tilde M}\Big(E_G~{\mathfrak M}_0(E_G,\vec p)- p_i~{\mathfrak M}_i(E_G,\vec p)\Big)
\label{long.5.3.1}
\end{eqnarray}
for $\tilde M =0$.  
Then the terms in eq. (\ref{long.5.1.2}) are expected to contribute, if the behavior of the expression
in eq. (\ref{long.5.3.1})
is like  $\tilde M^{-1}$  (as in matrix elements with unphysical modes). Otherwise they yield vanishing 
products (as for matrix elements  with only physical modes). 
\par
One can approach the problem of the decoupling of the longitudinal polarization in a
different but nevertheless interesting way:
by considering the sum of the contribution of eq. (\ref{long.5.3}) and that
of the spin zero part 
\begin{eqnarray}
\frac{d^3p}{2E_G}\,\,\,\frac{1}{{\tilde M}\,^2} \,\,\,{\mathfrak M}^*_\mu \,\,\,p^{\mu} p^{\nu} \,\,\,{\mathfrak M}_\nu.
\label{long.5.4}
\end{eqnarray}
With an algebra similar to the one in eq. (\ref{long.5.3}) one concludes that in the limit
of ${\tilde M}=0$ the contributions of the longitudinal polarization and of the
spin zero cancel if only physical states are present (i.e. eq. (\ref{long.4}) is valid).
A scholarly example in Appendix \ref{app:free}, based on the free fields, illustrates some of
the features of the limit $\tilde M =0$, discussed in the present Section.
\par
Now we compare the two quite different situations present in  the abelian and the nonabelian 
gauge theories.
\par\noindent
The Lagrange multiplier $b$, used for the gauge fixing,
decouples from the physical modes, as can be seen by using the STI. In the abelian case this yields
eq. (\ref{long.4}).
Consequently the contribution of the longitudinal polarization can be replaced
according to eq. (\ref{long.3}) in the zero mass limit, since the 
the expression in the second term of eq. (\ref{long.5}) will never multiply a $\tilde M^{-1}$ 
factor. 
\par
In nonabelian gauge theories the decoupling does not happen. In fact the decoupling
of the Lagrange multipliers does not bring to the eq. (\ref{long.4}), but instead
to a relation involving the Goldstone bosons as discussed later on.
This is the source
of many problems. For a single external gauge particle with longitudinal
polarization the replacement (\ref{long.3}) does yield the correct
result. However already for two gauge particles with longitudinal
polarization the replacement (\ref{long.3}) might gives results that depend
on the gauge or on the order of the replacements. The first
replacement gives no problems because all other particles are
physical. After the first replacement the $ {\cal O}(\frac{{\tilde M}}{E})$ term 
in eq. (\ref{long.3}) might yield non zero contributions
involving the Faddeev-Popov ghosts, since the spin zero
part of the gauge boson ($\epsilon_\mu \simeq p_\mu, p^2= {\tilde M}_G\,^2$)
is an unphysical mode.
\par 
This fact has further unpleasant consequences. For instance in the proof of physical
unitarity the sum over final states is only on physical modes. 
After the replacement (\ref{long.3}) (a physical mode by an unphysical one) this
necessary property is lost, if no further condition is introduced
to cancel out the spurious terms.
\section{Metamorphosis in the Higgs Mechanism Scenario}
\label{sec:cut}
 For ${\tilde M}=0$ only the transverse polarizations are physical,
thus there is a problem in the limit. In the massless
case the two unphysical modes of the vector fields
conspire with the Faddeev-Popov ghosts in order to cancel
out in the cutting rule, i.e. in the equation of perturbative
unitarity. Instead, for every finite value of ${\tilde M}$ the net balance to zero
involves the spin zero part of the vector mesons, the Goldstone bosons
and the Faddeev-Popov ghosts. Therefore
if the longitudinal polarization does not decouple,
unitarity is violated in the limit. This has been noticed a long time ago
\cite{Slavnov:1970tk}, \cite{Curci:1976kh}.
\par
The conceptual difficulty of the limit disappears if we accept
a scenario where the longitudinal mode transforms into the 
former Goldstone boson for zero vector meson mass. In fact the
Goldstone boson field describes  a physical mode at $\tilde M=0$.
This scenario is in agreement with the discussion of Section \ref{sec:long}
about the impossibility to distinguish the modes at very high energy.
\par
The metamorphosis scenario has a further advantage for practical
calculations: one can use the limit theory in order to evaluate
the amplitudes involving the longitudinal modes. 
This statement is very close to the CLTCG theorem which relates
the S-matrix elements of the longitudinal modes to those of the 
Goldstone boson. This advantage, however,
is limited by the infrared divergences, that eventually will emerge
(for instance in self-energies). Our scenario provides a more flexible setup, where the objections
on the zero mass limit are removed (metamorphosis versus decoupling) and a
proper use of the theory is established (only at $M\not =0$ we have a {\sl bona fide}
theory and the S-matrix elements at $M=0$ can be evaluated by using
$M$ as an infrared regulator).
In the next Section we give some comments about the CLTCG theorem.
\section{Comments on the CLTCG Theorem} 
\label{sec:equiv}
In this Section we provide the general formulation of the CLTCG
Theorem in any covariant 't Hooft gauge.
\par
Let $|\vec p L\rangle$ denotes an asymptotic state longitudinally polarized
with momentum $\vec p$. Since it is a physical state then
it must be annihilated by the operator $F$ that generates the BRST
transformations on the fields of the nonabelian gauge theory (internal index
is not displayed) \cite{Curci:1976yb}.
\begin{eqnarray}
F |\vec p L\rangle =0.
\label{cut.1}
\end{eqnarray}
However the state is also represented by any element of the equivalent
class made of vectors like
\begin{eqnarray}
 |\vec p L\rangle + F |X\rangle,
\label{cut.2}
\end{eqnarray}
where $X$ is an arbitrary state.
Due to the nilpotency of $F$, eq. (\ref{cut.1}) is still valid
\begin{eqnarray}
F\Bigl ( |\vec p L\rangle + F |X\rangle \Bigr )=0.
\label{cut.3}
\end{eqnarray}
By the standard proof of physical unitarity, the states $|\vec p L\rangle$ and 
$ |\vec p L\rangle + F |X\rangle$ have the same $S-$ matrix elements.
\par
We shall use this freedom in describing the physical modes of the gauge
fields in order to evaluate the behavior for ${\tilde M}\to 0$, without using
the replacement (\ref{long.3}) and encountering  some pitfalls. 
The recipe is the following: any longitudinal mode is replaced
by
\begin{eqnarray}
 |\vec p L\rangle +\frac{1}{{\tilde M}}~ F |\vec p \,\,\bar c\rangle,
\label{cut.4}
\end{eqnarray}
where $|\vec p \,\,\bar c\rangle$ is a single-mode anti-ghost state
and the relative weight is chosen in order to reproduce eq. (\ref{basic.11.3}),
when the wave functions are exhibited by the reduction formulas. 
We construct the in- and out-states in the Fock space by using
the recipe in eq. (\ref{cut.4}).
\par
The $S-$matrix element
for the longitudinal mode is constructed by using the amputated
connected Green function defined by 
\footnote{$\psi$ is an irreducible set of fields.
Throughout the paper we use the notation
\begin{eqnarray}
W_{ A^*_{a\mu}\dots}\equiv  \frac{\delta^n W}{\delta A^*_{a\mu} \dots}
=  i^{n-1}\langle 0|T((D^\mu [A]c)_a\dots)|0\rangle_C
\label{app1.9}
\end{eqnarray}
for composite fields, while for elementary fields
\begin{eqnarray}
W_{ \underbrace{b_a\dots}_n}\equiv  i^{n-1}\langle 0|T(b_a\dots)|0\rangle_C.
\label{app1.10}
\end{eqnarray}
For the effective action we use a similar short notation
\begin{eqnarray}
\Gamma_X\equiv  \frac{\delta \Gamma}{\delta X}.
\label{app1.7}
\end{eqnarray}
}
\begin{eqnarray}
W_{{A_\mu}} =   \sum_{\psi} W_{A_\mu \psi} W_{\widehat{\psi}}
\label{cut.5}
\end{eqnarray}
The asymptotic states are described by the eigenvectors $\epsilon^{(r)}$ 
and eigenvalues $\lambda^{(r)}$
of the residuum matrix of the two-point connected function $-W(p)$
at the physical pole $p^2=m^2_r$. 
The construction of the $S-$matrix element where the state $|\vec p L\rangle$
appears as a factor in the final state proceed via the usual procedure
(wave function renormalization factor and internal indexes are omitted)
\begin{eqnarray}
S_{\vec p L\cdots} \simeq {\epsilon_{L\mu}}(\vec p)
~i~W_{\widehat{A_\mu(p)}***}\Big|_{p^2={\tilde M}^2} \,  .
\label{cut.5.1}
\end{eqnarray}
With these notations the
residuum of the $b-$ field (the Lagrange multiplier of the 't Hooft
gauge) yields (see eq. (\ref{basic.8}))
\begin{eqnarray}
\lim_{p^2={\tilde M}_G^2}(p^2-{\tilde M}^2_G) W_{b(p) ***}=\biggl(
i\frac{p^\nu\Gamma_{\phi A^\nu}}{\Gamma_{\phi\phi}}W_{\widehat{\phi(p)} ***}
+ip^\mu W_{\widehat{A^\mu(p)} ***}\biggr)\Big |_{p^2={\tilde M}_G^2}
=0,
\label{cut.6}
\end{eqnarray}
where $***$ denotes more $b$ and  physical mode insertions. It's worth noticing
that from eq. (\ref{basic.99})
\begin{eqnarray}
i\frac{p^\nu\Gamma_{\phi A^\nu}}{\Gamma_{\phi\phi}}\Big |_{p^2={\tilde M}_G\,^2}
= \frac{\xi}{ M} {\tilde M}_G\,^2
\label{cut.6.1}
\end{eqnarray}
and at the tree level
\begin{eqnarray}
i\frac{p^\nu\Gamma_{\phi A^\nu}}{\Gamma_{\phi\phi}}\Big |_{p^2= M_G^2}
=  M.
\label{cut.6.2}
\end{eqnarray}
In order to reproduce the pattern of eq. (\ref{long.2}), 
the Feynman amplitude is replaced, according to  (\ref{cut.4}),
\begin{eqnarray}
\epsilon_{L\mu }(p)
W_{\widehat{A_\mu(p)}***}\Big|_{p^2={\tilde M}\,^2} =
\lim_{p^2={\tilde M}\,^2}  \epsilon_{L\mu} W_{\widehat{A_\mu(p)} ***} 
+i\lim_{{p^2={\tilde M}_G\,^2}}(p^2-{\tilde M}_G\,^2)\frac{1}{{\tilde M}}W_{b(p) ***}
\label{cut.7}
\end{eqnarray}
and by using eq. (\ref{cut.6})
\begin{eqnarray}&&
\epsilon_{L\mu }(p)
W_{\widehat{A_\mu(p)}***}\Big|_{p^2={\tilde M}\,^2} 
= 
\lim_{p^2={\tilde M}\,^2}  \epsilon_{L\mu} W_{\widehat{A_\mu(p)} ***} -\lim_{p^2={\tilde M}_G\,^2} \frac{1}{{\tilde M}}
\Bigl(p^\mu W_{\widehat{A^\mu(p)} ***} 
\nonumber\\&&
+
\frac{p^\nu\Gamma_{\phi A^\nu}}{\Gamma_{\phi\phi}}W_{\widehat{\phi(p)} ***}
\Bigr).
\label{cut.7.1}
\end{eqnarray}
The above replacement (\ref{cut.4}) may be repeated for every external gauge line
with longitudinal polarization, since on both terms in eq. (\ref{cut.7}) this replacement
is allowed. In fact the external legs are either physical states or $b$-lines and 
therefore the STI (\ref{app2.7}) guarantees the validity of eq. (\ref{cut.6}). 
\par
In the limit $M=0$
the first two terms in the RHS of eq. (\ref{cut.7.1}) cancel out
\begin{eqnarray}
\lim_{p^2=\tilde M^2}  \epsilon_L^\mu W_{\widehat{A_\mu(p)} ***} 
-\lim_{p^2=\tilde M^2_G} \frac{1}{\tilde M}
p^\mu W_{\widehat{A^\mu(p)} ***}= 0 + {\cal O}(\frac{\tilde M}{E}).
\label{cut.8}
\end{eqnarray}
Thus finally we get
\begin{eqnarray}&&
\epsilon_L^{\mu }(p)
W_{\widehat{A_\mu(p)}***}\Big|_{p^2={\tilde M}\,^2} 
=  
i~\frac{ p^\nu\Gamma_{\phi A^\nu}}{\tilde M\Gamma_{\phi\phi}}~W_{\widehat{\phi(p)} ***}
\Big |_{p^2=\tilde M^2_G}
+ {\cal O}(\tilde M).
\label{cut.9}
\end{eqnarray}
The procedure can be repeated for every vector boson in the longitudinal mode. The feared
occurrence of cross terms ${\cal O}(M) \times \frac{1}{M}$ vanishes since all external modes
are either physical or $b-$insertions.
\par 
Few comments are in order on our proof of the CLTCG theorem in eq. (\ref{cut.9}).
\begin{enumerate}
\item The exact knowledge of the two-point functions in the unphysical sector, 
as displayed in Sec. \ref{sec:ptpf},
allows the correct formulation of the CLTCG theorem at any number of loops. The quantities needed
are $\Gamma_{\phi\phi}$ and $p^\mu \Gamma_{\phi A^\mu}$.
\item The CLTCG theorem as in eq. (\ref{cut.9}) concerns the amputated connected amplitudes.
In order to formulate the theorem for the $S-$matrix elements one has to 
introduce the wave-function normalization of the asymptotic states. 
In the limit $M=0$ the normalization of the longitudinal modes equals that
of the Goldstone boson, since $W_L$ approaches the free-field value,
as displayed in eq. (\ref{ptpf.31.3}).
Here we are not going into further details on this problem.
\item After we introduce the necessary wave function renormalization factor in the LHS of
eq. (\ref{cut.9}) the $S$-matrix elements are gauge invariant. 
This is valid for any finite value of $\tilde M$.
Thus also the limit, when it exists, is gauge invariant.
The property has been verified  in explicit calculations \cite{Denner:1997kq}.
\item For generic $S$-matrix elements the limit of zero mass is expected
to be infrared divergent. Then eq. (\ref{cut.9}) is of no use. However
one might consider quantities that are infrared finite as, for instance, 
some transition probabilities. On those quantities
the theorem in eq. (\ref{cut.9}) might apply. A further possibility is to
work in generic dimension D, whenever it is possible.
\item Physical unitarity is valid at every value of ${\tilde M}$.
In case of the Higgs mechanism scenario we get the hint for 
considering a metamorphosis of the longitudinally polarized
vector meson into a massless scalar at ${\tilde M}=0$.
\item
Within the Higgs mechanism scenario the limit ${\tilde M}=0$
can be performed on the perturbative expansion by providing a consistent
picture of the metamorphosis of the longitudinal mode. Thus eq. (\ref{cut.9})
can be read in the other way around: the ${\tilde M}\not = 0$ theory provides
a doable infrared regulator for the $v=0$ theory (symmetric phase), when on-shell
amplitudes are needed.
\end{enumerate}
The CLTCG theorem as in eq. (\ref{cut.9}) provides a tool for solving the
problem associated to the longitudinal mode that does not decouple from
physical states in the zero mass limit. The scenario of a metamorphosis
of this mode into the Goldstone boson field, which can be consistently
taken as a physical mode at zero mass, looks very promising for satisfying
perturbative unitarity and the set of relations derived from the 
STI, gauge fixing equation and anti-ghost equation.
This setting is compatible with the picture of a symmetry restoration
through the limit $v=0$ on the perturbative series of the effective
action. It is tempting to argue that this setting allows the limit $v=0$ in a nonabelian
gauge theory coupled with scalars (the former, i.e. $v\not=  0$, Higgs field),
i.e. the generating functionals are continuous in $v=0$.
Such dynamical theory remains non asymptotically free, according to
the classification of Refs. \cite{Gross:1973ju}-\cite{Politzer:1973fx}.
\section{Zero Mass Limit with a St\"uckelberg Gauge Invariant Term}
\label{sec:stuck}
The equations used to support the metamorphosis
scenario are still valid in the case of a Yang-Mills theory 
with mass {\sl \`a la} St\"uckelberg. In particular one has
the same STI, gauge-fixing equation and 
anti-ghost equation. However the use of eq. (\ref{cut.9}) is 
now in question: although the formal derivation is the same,
the non-existence of a zero mass limit in the loop
expansion makes the CLTCG theorem inapplicable. 
\par
For small $M$ many terms of perturbative expansion have
singular $M^{-1}$ behavior.
In fact, in the nonlinear theories the perturbative expansion in 
the loop number works for momenta small with respect to $M$.
It is not known how this region is connected to the
one around $M=0$. Consequently the extension of the CLTCG theorem 
to the massless limit becomes questionable.
\par
The study of the zero-mass region implies 
a typical strong-coupling limit: the $M^{-2}$ factor
in the $\phi$ propagator is responsible for the 
presence of many divergent terms in the perturbative 
expansion. This means that one cannot explore the
$M=0$ region by starting from the series expansion in the number 
of loops.
\par
One can make an educated guess on the small mass behavior
of the theory on the basis of some na{\"\i}ve considerations. 
\par\noindent
i) $M$ controls in some way the spontaneous breakdown of 
the global $SU(2)_L\otimes SU(2)_R$ transformations. In fact the St\"uckelberg
mass term is the source of the interaction of $\phi$ with
the rest of the fields. The order parameter $\langle \phi_0\rangle$
for the SBS has no effect at  $M=0$. Thus a limit theory is expected
to be symmetric. 
\par\noindent
ii) BRST properties of the asymptotic fields (if they can be defined)
indicate that the fields  $\vec\phi$ remain unphysical for
any  value of  $M$, since the vacuum expectation value
of  $\phi_0$ is expected to remain non-zero. 
\par\noindent
iii) 
One might consider a resummation of the series, by performing
first the integration over the $SU(2)$ group, thus taking
$A_\mu$ as an external source. One gets an expansion in 
powers of $M$ where the coefficients are invariant under local gauge transformations.
In this setting the path integral on the fields $\vec\phi$ is performed
on a lattice, with spacing $a$,
\begin{eqnarray}&&
\int {\cal D}[\phi] \exp 
\Big(\int_E d^4 x  \frac{M^2}{2g^2} (A_{a\mu} - F_{a\mu})^2 \Big)
\nonumber\\&&
\simeq
\int {\cal D}\Omega
\exp \sum_{x\mu}\Biggl [
 \frac{ 2 M^2}{g^2}a^{2} {\mathfrak Re}
\sum_{x\mu}Tr\Bigl\{ \Omega(x)^\dagger U(x,\mu)\Omega(x+\mu)-1\Bigr\} \Biggr],
\label{stuck.1}
\end{eqnarray}
where link variable is function solely of the classical field
$A_\mu$, the remaining integration variable in the final expression
of the generating functional. The path integral integration is over
a compact set for each site, therefore we can expand in powers of $M$
\begin{eqnarray}&&
\frac{1}{\int \prod_x {\cal D}[\Omega(x)]}
\int \prod_x {\cal D}[\Omega(x)]\Biggl[1
+\frac{1}{2} \Bigl (\frac{\beta M^2
  a^2}{2}\Bigr)^2 \Bigl ( \sum_{x\mu} Tr \{\Omega^\dagger(x)U(x,\mu)\Omega(x+\mu)\}\Bigr)^2
\nonumber\\&&
+\frac{1}{4!}(\frac{2 M^2 a^2}{g^2})^4\prod_{j=1}^4
\Bigl(
\sum_{x_j\mu_j}Tr\Bigl\{ \Omega(x_j)^\dagger U(x_j,\mu_j)\Omega(x_j+\mu_j)\Bigr\}
\Bigr)\Biggr]
\nonumber\\&&
=1+ \frac{1}{2} \Bigl (\frac{\beta M^2
  a^2}{2}\Bigr)^2 DN + \frac{1}{8} \Bigl (\frac{\beta M^2
  a^2}{2}\Bigr)^4 (DN)^2 + \frac{1}{4}(\frac{ M^2 a^2}{g^2})^4\sum_\Box Tr \Bigl\{ U_\Box \Bigr\},
\label{stuck.1.1}
\end{eqnarray}
where $U_\Box$ is the $SU(2)$ matrix associated to the plaquette $\Box$ and $DN$ is 
the number of degrees of freedom of the gauge bosons times the number
of sites.
Subsequently we take into account the integration over the link variables.
Finally we take the logarithm of the partition function 
\begin{eqnarray}
\!\!\! \ln Z = \ln Z_0 +  \frac{1}{2} \Bigl (\frac{\beta M^2
  a^2}{2}\Bigr)^2 DN  +\frac{1}{4}\Bigl(\frac{\beta M^2 a^2}{2}\Bigr)^4 \Big\langle \sum_\Box 
  Tr\{U_\Box \}\Big \rangle \Big |_{M^2a^2=0},
\label{stc.4.1}
\end{eqnarray}
where $Z_0$ is the partition function of the massless theory.
Thus the final result is a  Yang-Mills theory with local insertions.
\par
There is another point in favor of this guess and it
comes from eq. (\ref{cut.9}). From completely general consideration
(i.e. no approximations are needed) we have argued that the
limit of zero mass can be performed by keeping BRST invariance,
gauge invariance of the $S$-matrix and perturbative unitarity.
If the limit implied by the CLTCG theorem exists
in the form of a well-defined local theory represented
on a Fock space of asymptotic fields, then both the longitudinal
polarization mode and the Goldstone boson must decouple in
the limiting region. If not, perturbative unitarity is
violated in default of cancellation of the unphysical modes.
BRST transformations on the asymptotic fields show that
$\vec\phi$ remains an unphysical mode, if not decoupled.
Therefore   the metamorphosis of a physical mode
(longitudinal polarization) into an unphysical mode (the Goldstone
boson) can not occur.
This is
an educated guess saying that the longitudinal polarization
mode decouples in the zero-mass limit.
\par
The resulting massless Yang-Mills theory is supposed to describe 
events where the mass is negligible (with respect to energies,
momentum transfers and any other dimensionful quantity). 
This point should be made clear:
no confinement is implied of any  sort.
\par
These arguments indicate that Yang-Mills with mass
generated by the Higgs mechanism or introduced by the
the St\"uckelberg term might be compared on phenomenological
ground since at very high energy the number of degrees of freedom
are different. Thus the two theories can be tested not
only by the direct detection of the Higgs boson but also
by this new very important difference in processes
at high energy.
\section{An Example}
\label{sec:exam}
The difference between nonabelian \underline{massive} 
gauge theories with Higgs mechanism and with non linear 
realization can be shown in many realms. The one-loop corrections
in the two theories have been discussed and analytically evaluated
in Ref. \cite{Bettinelli:2007cy}.
Moreover the two models  show marked differences in the celebrated
processes $WW$, $WZ$ and $ZZ$ elastic scattering. 
In fact, according
to the previously presented arguments,  the limit $M=0$
of the nonlinear theory
is a pure nonabelian gauge model without Higgs scalars and 
vector meson
longitudinal polarizations.
\par
In the present Section we  consider a process involving quarks or leptons
in order to illustrate the metamorphosis and its consequences.
In particular we focus on a process where no Higgs boson
is mediating, in order to have a signature which is not directly
connected to its existence 
\cite{Barger:1987nn}. 
\par
The present example is not intended as a quantitative argument
for future measurements. For the last scope one needs 
more involved calculations including, for instance, the loop contributions
of the top and the corrections due to the running of the constants.
This part of the research is not considered in the present
work.
\par
We consider the following process \cite{Altarelli:2000ye}
\begin{eqnarray}
 d + \bar u \to b + \bar t,
\label{exam.1}
\end{eqnarray}
where the intervening
quarks  might be easily replaced by  other constituents
(e.g. $l,\bar\nu$).
We follow the conventions of Ref. \cite{Nakamura:2010zzi}
for the fermion sector.
In case of Higgs mechanism we have a Drell-Yan process
mediated by $W^-$ and $\phi^-$. In the Landau gauge we have
(we consider only the $s-$channel graph)
\begin{eqnarray}&&
{\mathfrak M}(M) = \frac{g^2 V_{ud}V_{tb}^*}{2} \bar v_u \gamma_\mu\frac{1- \gamma_5}{2}u_d
\frac{g^{\mu\nu}-\frac{q^\mu q^\nu}{q^2}}{q^2-M^2}
\bar u_b \gamma_\nu \frac{1- \gamma_5}{2} v_{t}
\nonumber\\&&
+ V_{ud} V_{tb}^* \bar v_u \Big[f_u\frac{1- \gamma_5}{2}
- f_d \frac{1+ \gamma_5}{2}\Big]u_d \frac{1}{q^2}\bar u_b \Big[f_t\frac{1- \gamma_5}{2}
- f_b \frac{1+ \gamma_5}{2}\Big]v_t 
\label{exam.2}
\end{eqnarray}
with $M=gv$ and $\sqrt 2 f_x v= m_x$.
Unitarity is preserved on-shell; in fact at $q^2=M^2$ the
only pole is in the propagator of the $W^-$ with a residuum
that projects on the physical polarizations. Moreover there is no
pole at $q^2=0$, since the Goldstone boson cancels the spin zero
part of the vector meson. 
\par
Eq. (\ref{exam.2}) shows in detail what happens in the limit $v=0$:
we perform the limit in two different ways. First we add the contributions
of the Goldstone part to the vector meson propagator to obtain the unitary
gauge amplitude. On that amplitude  the limit is performed to discover
that the longitudinal polarizations yield a finite result. Second
the limit is taken separately on the two terms of eq. (\ref{exam.2}).
The gauge term $q^\mu q^\nu$ of the $W$-propagator vanishes in the limit, while the
Goldstone contribution survives to match the longitudinal polarization's
of the previous limit procedure, as in 
the mechanism described in eq. (\ref{long.5.1.2}).
the $q^\mu q^\nu$ term in the $W$-propagator vanishes via Dirac
equation, while the ``Goldstone'' field contribution survives, as in 
the mechanism described in eq. (\ref{long.5.1.2}).
This exemplifies the metamorphosis of the longitudinal polarization into
the physical massless scalar mode, originally associated
to the unphysical Goldstone boson for $v\not =0$.
\par \noindent
By taking the limit in the first fashion,
for every value of $M$ the two terms add to
\begin{eqnarray}
{\mathfrak M}(M) = \frac{g^2 V_{ud}V_{tb}^*}{2} \bar v_u \gamma_\mu\frac{1- \gamma_5}{2}u_d
\frac{g^{\mu\nu}-\frac{q^\mu q^\nu}{M^2}}{q^2-M^2}
\bar u_b \gamma_\nu \frac{1- \gamma_5}{2} v_{t}.
\label{exam.3}
\end{eqnarray}
The $M^{-2}$ term does survive in the limit of zero mass, since the $q^\mu q^\nu$
produces a quadratic term in the quark or lepton masses and therefore the $v^2$
dependence disappears in the ratio. 
While in the second way we take the $v=0$ limit on the
longitudinal part of the propagator in the Landau gauge ($q^\mu q^\nu/q^2$):
the result is zero. But the Goldstone contribution is finite and adds to the total amplitude
in eq. (\ref{exam.3}) taken at $v^2=0$.
\par
In the nonlinear theory, according to our guess, 
such terms are not present and the $W$-propagator in the Landau gauge is as usual
\begin{eqnarray}
-i
\frac{g^{\mu\nu}-\frac{q^\mu q^\nu}{q^2}}{q^2},
\label{exam.3.1}
\end{eqnarray}
where the $\frac{q^\mu q^\nu}{q^2}$ vanishes on the quark and lepton chiral currents, 
since the $M=0$ limit has been already taken.
Finally, the difference between the two scenarios (linear versus
nonlinear) are traced simply by the presence of a $\frac{1}{M^2}$ factor.
\par
Now we look whether the difference is of some phenomenological relevance,
in order to show the origin of the difficulty to find a measurable signature.
The square modulus of the amplitude (\ref{exam.3}) summed over the polarizations
of the incoming and outgoing particles yields
\begin{eqnarray}&&
\sum_{\rm POL}|{\mathfrak M}(M)|^2
= \frac{1}{(q^2-M^2)^2}\Bigg\{
16 (p_bp_u)( p_tp_d)
\nonumber\\&&
- \frac{8}{M^2}\Big[
m^2_t m^2_u (p_bp_d)+m^2_t m^2_d (p_bp_u)+
m^2_b m^2_u (p_tp_d)+m^2_b m^2_d (p_tp_u)
\Big]
\nonumber\\&&
+\frac{4}{M^4}\Big[ 2 m^2_t m^2_b+ ( m^2_b+  m^2_t)(p_bp_t)  \Big]
\Big[ 2 m_u^2 m_d^2+ (m_u^2+ m_d^2) (p_up_d) \Big]
\Bigg\}.
\label{exam.8}
\end{eqnarray}
It is clear that the $M^{-2}$ and $M^{-4}$ 
terms are negligible and therefore one cannot
discriminate the linear model (with Higgs boson) from the nonlinear one (without Higgs boson) in this process.
\section{Conclusions}
\label{sec:concl}
We consider the massive Yang-Mills theory
at very high momenta both in the case of a Higgs mechanism
generated mass and of a St\"uckelberg mass term. 
The kinematical set up is reproduced by the $M=0$ limit,
by assuming that only one energy scale is present in the
physical process. In this limit the number of degrees of
freedom of vector mesons changes. This fact poses a problem
for unitarity, since the longitudinal modes do not decouple
in nonabelian gauge theories for $M=0$.
\par
In the first case we suggest the metamorphosis of
the longitudinal modes into the Goldstone scalar bosons 
when the limit $M=0$ is taken. 
The scenario is supported
by the CLTCG theorem. {\sl In passing} we present some improvements
on the CLTCG theorem.  According to this proposal the unitarity
equation is consistently satisfied both for $M\not =0$
and $M=0$: no mismatch of degrees of freedom shows
up and moreover the symmetric limit  $v=0$ looks
smooth. The limit theory consists of a massless gauge
Yang-Mills in interaction with a doublet of 
physical complex scalar fields 
(the Higgs and the Goldstone bosons).
The theory is expected to be non-asymptotically free.
\par
In the St\"uckelberg mass case the limit $M=0$ on
the perturbative series is not possible due to very singular
terms. We suggest that the perturbative region is separated
from the $M\sim 0$ region by some singularity line between 
different phases. We envisage the scenario where at $M\sim 0$
both the Goldstone bosons and the longitudinal modes decouple
and the theory is realized in a confined phase, typical of
a massless gauge theory. However the properties of 
the confinement phase are relevant  only
for extremely high momenta ($M\sim 0$) processes (e.g. asymptotic freedom) and not
for low energy states.
\par
The conjecture establishes a ground for developing experimental tests
capable to distinguish a linearly (Higgs formalism)- from a nonlinearly 
(St\"uckelberg mass)-realized nonabelian massive gauge theory. 
However this aspect of the work  is beyond the scope of the present
paper.
\par
In the present paper we use the covariant 't Hooft
gauge. Several exact results are derived for the two-point
functions in the  unphysical sector. The limit at $M=0$ 
of these two-point functions is evaluated in the Higgs mechanism.
This limit tells that  the equivalence theorem (CLTCG) in its tree-level formulation
is not modified by loop corrections.
\section*{Acknowledgments}
I am honored to thank the warm hospitality of the
Center for Theoretical Physics at MIT, Massachusetts, where 
the present work  has been done.
This work is supported by U.S. Department of Energy (D.O.E.) under cooperative research 
agreement DE-FG0205ER41360.
\appendix
\section{Free Field Example }
\label{app:free}
In this Appendix we evaluate the contributions of
the longitudinal mode and of the spin zero part of the vector meson to
the unitarity sum in the free case. 
Consider the {\sl free} vector meson propagator in the generic $\xi$ gauge
\begin{eqnarray}
-\frac{g_{\mu\nu}-\frac{p_\mu p_\nu}{p^2}}{p^2-M^2} - \frac{1}{\xi}\frac{1}{p^2-\frac{M^2}{\xi}}
\frac{p_\mu p_\nu}{p^2}.
\label{tpf.65}
\end{eqnarray}
We use the residua of the poles both for $p^2=M^2$ and $p^2= \frac{M^2}{\xi}$
and at the end we take  the limit $M^2=0$. The contribution of the longitudinal polarization
on the pole $p^2=M^2$ as in eq. (\ref{long.1}) is
\begin{eqnarray}
\frac{1}{M^2}
\left (
\begin{array}{ll} p^2  &E p_j   \\
E p_i & \frac{E^2}{p^2}p_i p_j 
\end{array}
\right)
\label{tpf.66}
\end{eqnarray}
while the spin zero at $p^2= \frac{M^2}{\xi}$
\begin{eqnarray}
-\frac{1}{M^2}
\left (
\begin{array}{ll} E_{\scriptscriptstyle G}^2  &E_{\scriptscriptstyle G}p_j  \\
E_{\scriptscriptstyle G} p_i &p_i p_j 
\end{array}
\right).
\label{tpf.67}
\end{eqnarray}
The contribution of the two transverse polarizations is
\begin{eqnarray}
\left (
\begin{array}{ll}0&0  \\
0 & \delta_{ij}-\frac{1}{p^2}p_i p_j 
\end{array}
\right).
\label{tpf.68}
\end{eqnarray}
The expression in eq. (\ref{tpf.66}) is
multiplied by some quantity $M^{\mu\nu}(E,  \vec p)$, while that in
eq. (\ref{tpf.67}) by  $M^{\mu\nu}(E_{\scriptscriptstyle G},  \vec
p)$. 
\par\noindent
Now we add the contributions with  
the front factors $1/E$ and $1/E_{\scriptscriptstyle G}$ originating from the
Lorentz invariant measure . Thus we have to add the terms
\begin{eqnarray}
\frac{1}{2EM^2}
\left ( p^2 M^{00}(E,  \vec p)+E p_j M^{0j}(E,  \vec p) +
E p_i M^{i0}(E,  \vec p) + \frac{E^2}{p^2}p_i p_j  M^{ij}(E,  \vec p)
\right)
\label{tpf.70}
\end{eqnarray}
and
\begin{eqnarray}&&
-\frac{1}{2E_{\scriptscriptstyle G} M^2}
\left ( E_{\scriptscriptstyle G}^2 M^{00}(E_{\scriptscriptstyle G},  \vec p)
+E_{\scriptscriptstyle G} p_j M^{0j}(E_{\scriptscriptstyle G},  \vec p) +
E_{\scriptscriptstyle G} p_i M^{i0}(E_{\scriptscriptstyle G},  \vec p) 
+ p_i p_j  M^{ij}(E_{\scriptscriptstyle G},  \vec p)
\right)
\nonumber\\&&
=
-\frac{1}{2EM^2}(1-\frac{E_{\scriptscriptstyle G}-E}{E} )
\Big [\left(p^2+\frac{M^2}{\xi}\right)\Big( M^{00}(E)+(E_{\scriptscriptstyle
  G}-E)\frac{\partial}{\partial E} M^{00}(E)\Big)
\nonumber\\&&
+ p_j \Big(E  M^{0j}(E)+(E_{\scriptscriptstyle G}-E)(M^{0j}(E)
+E\frac{\partial}{\partial E} M^{0j}(E)\Big)
\nonumber\\&&
+ p_i \Big(E  M^{i0}(E)+(E_{\scriptscriptstyle G}-E)(M^{i0}(E)
+E\frac{\partial}{\partial E} M^{i0}(E)\Big)
\nonumber\\&&
+ p_i p_j\Big ( M^{ij}(E)
+(E_{\scriptscriptstyle G}-E)\frac{\partial}{\partial E} M^{ij}(E)\Big)
\Big ]
\label{tpf.71}
\end{eqnarray}
Now we have
\begin{eqnarray}
E_{\scriptscriptstyle G}-E
\simeq -  M^2\frac{1-\frac{1}{\xi} }{2p}
\label{tpf.72}
\end{eqnarray}
and therefore we get
\begin{eqnarray}&&
\frac{1}{2p}\Bigg[
\frac{p}{2}(1-\frac{1}{\xi})\frac{\partial}{\partial E} M^{00}(E)
-\frac{1}{\xi} M^{00}(E)
\nonumber\\&&
+\frac{p_j}{2p}(1-\frac{1}{\xi})\Big(M^{0j}(E)
+E\frac{\partial}{\partial E} M^{0j}(E)\Big)
\nonumber\\&&
+\frac{p_i}{2p}(1-\frac{1}{\xi})\Big(M^{i0}(E)
+E\frac{\partial}{\partial E} M^{i0}(E)\Big)
\nonumber\\&&
+ \frac{p_ip_j}{p^2}  M^{ij}(E) 
+ \frac{p_ip_j}{2p}(1-\frac{1}{\xi})\frac{\partial}{\partial E} M^{ij}(E)\Bigg]
\nonumber\\&&
-\frac{1}{4p^3}(1-\frac{1}{\xi})\Bigg\{p^2 M^{00}+pp_jM^{0j}+p p_i M^{i0} +p_ip_jM^{ij}
\Bigg\}
\label{tpf.73}
\end{eqnarray}
If we add the transverse part (\ref{tpf.68}) we get
\begin{eqnarray}&&
\frac{1}{2p}\Bigg[
\frac{p}{2}(1-\frac{1}{\xi})\frac{\partial}{\partial E} M^{00}(E)
-\frac{1}{\xi} M^{00}(E)
\nonumber\\&&
+\frac{p_j}{2p}(1-\frac{1}{\xi})\Big(M^{0j}(E)
+E\frac{\partial}{\partial E} M^{0j}(E)\Big)
\nonumber\\&&
+\frac{p_i}{2p}(1-\frac{1}{\xi})\Big(M^{i0}(E)
+E\frac{\partial}{\partial E} M^{i0}(E)\Big)
\nonumber\\&&
+ \delta_{ij} M^{ij}(E) 
+ \frac{p_ip_j}{2p}(1-\frac{1}{\xi})\frac{\partial}{\partial E} M^{ij}(E)
\nonumber\\&&
-\frac{1}{2p^2}(1-\frac{1}{\xi})\Big\{p^2 M^{00}+pp_jM^{0j}+p p_i M^{i0} +p_ip_jM^{ij}
\Big\}
\Bigg]
\label{tpf.74}
\end{eqnarray}
that can be written
\begin{eqnarray}&&
\frac{1}{2p}\Bigg(
-g_{\mu\mu} M^{\mu\nu}(E) +(1-\frac{1}{\xi})
\Bigg[ M^{00}(E)
+\frac{p}{2}\frac{\partial}{\partial E} M^{00}(E)
\nonumber\\&&
+\frac{p_j}{2p}\Big(M^{0j}(E)
+E\frac{\partial}{\partial E} M^{0j}(E)\Big)
\nonumber\\&&
+\frac{p_i}{2p}\Big(M^{i0}(E)
+E\frac{\partial}{\partial E} M^{i0}(E)\Big)
+ \frac{p_ip_j}{2p}\frac{\partial}{\partial E}
M^{ij}(E)
\nonumber\\&&
-\frac{1}{2p^2}\Big\{p^2 M^{00}+pp_jM^{0j}+p p_i M^{i0} +p_ip_jM^{ij}
\Big\}
\Bigg]\Bigg |_{E=|\vec p|} \Bigg)
\nonumber\\&&
=
\frac{1}{2p}\Bigg(
-g_{\mu\mu} M^{\mu\nu}(E) +(1-\frac{1}{\xi})
\Bigg[\frac{1}{2} M^{00}(E) -\frac{p_ip_j}{2p^2}M^{ij}
+\frac{p}{2}\frac{\partial}{\partial E} M^{00}(E)
\nonumber\\&&
+\frac{p_j}{2}\frac{\partial}{\partial E} M^{0j}(E)
+\frac{p_i}{2}\frac{\partial}{\partial E} M^{i0}(E)
+ \frac{p_ip_j}{2p}\frac{\partial}{\partial E}
M^{ij}(E)
\Bigg]\Bigg |_{E=|\vec p|} \Bigg)
.
\label{tpf.75}
\end{eqnarray}
This agrees with the $M=0$ limit propagator
\begin{eqnarray}
-\frac{g_{\mu\nu}-(1-\frac{1}{\xi})\frac{p_\mu p_\nu}{p^2}}{p^2}.
\label{tpf.76}
\end{eqnarray}
In fact the positive frequency pole  gives the first term in eq. (\ref{tpf.75})
while the  double pole gives
\begin{eqnarray}&&
(1-\frac{1}{\xi})
\frac{\partial}{\partial p_0}\frac{p_\mu
  p_\nu}{(p_0+p)^2}M^{\mu\nu}(p_0) |_{p_0=|\vec p|}
\nonumber\\&&
=(1-\frac{1}{\xi})\Bigg[\frac{1}{2p}M^{00}(p)
-\frac{1}{4p}M^{00}(p)+\frac{1}{4}\frac{\partial}{\partial
  p_0}M^{00}(p)
\nonumber\\&&
+\frac{p_j}{4p^2} M^{0j}-\frac{p_j}{4p^2}M^{0j}+\frac{p_j}{4p}\frac{\partial}{\partial
  p_0}M^{0j}(p)
\nonumber\\&&
+\frac{p_i}{4p}\frac{\partial}{\partial
  p_0}M^{i0}(p) -\frac{p_ip_j}{4p^3}M^{ij}(p)+\frac{p_ip_j}{4p^2}\frac{\partial}{\partial
  p_0}M^{ij}(p)
\Bigg]
\nonumber\\&&
=\frac{1}{2p}(1-\frac{1}{\xi})\Bigg[
\frac{1}{2}M^{00}(p)+\frac{p}{2}\frac{\partial}{\partial
  p_0}M^{00}(p)
+\frac{p_j}{2}\frac{\partial}{\partial
  p_0}M^{0j}(p)
\nonumber\\&&
+\frac{p_i}{2}\frac{\partial}{\partial
  p_0}M^{i0}(p) -\frac{p_ip_j}{2p^2}M^{ij}(p)+\frac{p_ip_j}{2p}\frac{\partial}{\partial
  p_0}M^{ij}(p)
\Bigg]
.
\label{tpf.77}
\end{eqnarray}
The RHS of eq. (\ref{tpf.77}) is zero in presence of the conservation laws
$p^\mu M_{\mu\nu}= p^\nu M_{\mu\nu}=0$. 
\par\noindent
This Appendix shows that terms in eq. (\ref{long.5}), 
otherwise neglected on-shell, give essential
contributions for the limit $M=0$.
\section{BRST transformations and STI (Higgs)}
\label{app:app1}
We discuss the BRST transformations  for both models.
We derive the STI  and the
relations that are needed for the CLTCG theorem.
First we consider the case where the mass is generated 
by the Higgs mechanism.
\par
The necessity to maintain physical unitarity after
the introduction of a gauge fixing term requires
the use of the Faddeev-Popov ghosts. This is easily
done by imposing BRST invariance of the action
under the transformations
\begin{eqnarray}&&
{\mathfrak s} A_{a\mu} = (D_\mu [A]c)_a, \qquad 
{\mathfrak s} \phi_0 = -
\frac{1}{2} \phi_a c_a
 \nonumber \\&& 
{\mathfrak s} \phi_a = 
\frac{1}{2} \phi_0 c_a + \frac{1}{2} \epsilon_{abc} \phi_b c_c 
 \nonumber \\&& 
{\mathfrak s} \bar c_a = b_a , \qquad  {\mathfrak s}b_a = 0. 
\label{app1.1}
\end{eqnarray}
In the above equation $D_\mu[A]$ denotes the covariant derivative 
w.r.t. $A_{a\mu}$:
\begin{eqnarray}
(D_\mu[A])_{ac} = \delta_{ac} \partial_\mu + \epsilon_{abc} A_{b\mu} \, .
\label{app1.2}
\end{eqnarray}
The BRST transformation of $c_a$ then follows by nilpotency 
\begin{eqnarray}
{\mathfrak s} c_a = - \frac{1}{2} \epsilon_{abc} c_b c_c \, .
\label{app1.3}
\end{eqnarray}
%
%
Now we can easily obtain a BRST invariant action by
making invariant the gauge fixing term. This is achieved 
by using the nilpotency of $ {\mathfrak s}$. For the
generic covariant 't Hooft gauge we have
\begin{eqnarray}
S_{\rm H\,_{gf}} \to
S_{\rm  H\, _{GF}}  = \frac{\Lambda^{(D-4)}}{g^2} \int d^Dx \,
 {\mathfrak s} \,
\biggl[ \bar c_a\,\Big( \frac{b_a}{2\xi }+\frac{ M}{\xi }\phi_a +
 \partial_\mu A_a^\mu\Big )\biggr]. 
\label{app1.4}
\end{eqnarray}
The STI associated to the above BRST transformations, in the
notations of Batalin and Vilkovisky \cite{Batalin:1981jr,Batalin:1984jr}
can be easily derived. By introducing the external sources
\begin{eqnarray}
\int d^Dx \, (A_{a\mu}^* ~ {\mathfrak s} A_a^\mu 
+ \phi_a^* ~{\mathfrak s} \phi_a+ \phi_0^* ~{\mathfrak s} \phi_0 
 + c_a^* ~{\mathfrak s} c_a ) \, .
\label{app1.5}
\end{eqnarray}
for the 1-PI functional one gets
\begin{eqnarray}
&&
\int d^Dx \, \Big (
 \Gamma_{ A^*_{a\mu}}  \Gamma_{ A_a^\mu}
+
 \Gamma_{ \phi_a^*}  \Gamma_{ \phi_a}
+
 \Gamma_{ \phi_0^*}  \Gamma_{ \phi_0}
+ 
 \Gamma_{ c_a^*} \Gamma_{ c_a}
+ b_a  \Gamma_{ \bar c_a} 
\Big ) = 0 \, .
\label{app1.6}
\end{eqnarray}
While  for the generating functional of the connected amplitudes
one has
\begin{eqnarray}
&&
\int d^Dx \, \Big (
- W_{ A^*_{a\mu}}J_{a\mu} 
-
 W_{ \phi_a^*} K_a
-  W_{ \phi_0^*}K_0  
+ 
 W_{ c_a^*}\bar\eta_a
- W_{  b_a} \eta_a
\Big ) = 0\, . 
\label{app1.8}
\end{eqnarray}
%
The equation associated to the gauge fixing
gives
\begin{eqnarray}
\Gamma_{b_a}
=\Lambda_g \Big(
 \frac{b_a}{\xi }+\frac{ M}{\xi }\phi_a +
 \partial_\mu A_a^\mu \Big)
\label{app1.11}
\end{eqnarray}
\begin{eqnarray}
-J_{b_a}
=\Lambda_g \Big(
 \frac{1}{\xi }W_{b_a}+\frac{ M}{\xi }W_{\phi_a} +
 \partial_\mu W_{ A_a^\mu} \Big).
\label{app1.12}
\end{eqnarray}
The anti-ghost equation
\begin{eqnarray}
\Gamma_{\bar c_a}
=\Lambda_g \Big[
-\frac{ M}{2\xi }(c_a \phi_0 + \epsilon_{abc}c_c \phi_b)-
 \partial_\mu \left(D^\mu[A]c\right)_a \Big]
 =\Lambda_g \Big[
-\frac{ M}{\xi }\Gamma_{\phi_a^*}-
 \partial_\mu \Gamma_{A_{a\mu}^*}\Big]
\label{app1.13}
\end{eqnarray}
\begin{eqnarray}
\eta_a
=\Lambda_g \Big[
\frac{ M}{\xi }W_{\phi_a^*}+
 \partial_\mu W_{A_{a\mu}^*}\Big].
\label{app1.14}
\end{eqnarray}
%
\section{BRST transformations and STI (St\"uckelberg)}
\label{sec:st}
In the case where the mass of the Yang-Mills fields
comes from a St\"uckelberg term, the same BRST
apply as in eqs. (\ref{app1.1}) and (\ref{app1.3}).
Moreover the STI and the gauge fixing equation are akin to those
of Yang-Mills with Higgs mechanism.
Here we illustrate this property, which is not trivial
due to the fact that St\"uckelberg's mass term makes the theory
nonrenormalizable.
\par
Eq. (\ref{stck.1}) becomes
\begin{eqnarray}
S_{\rm S\,_{gf}} \to
S_{\rm  S\, _{GF}}  = \frac{\Lambda^{(D-4)}}{g^2} \int d^Dx \,
 {\mathfrak s} \,
\biggl[ \bar c_a\,\Big( \frac{b_a}{2\xi }+2\frac{ M^2}{\xi }\phi_a +
 \partial_\mu A_a^\mu\Big )\biggr]. 
\label{app2.1}
\end{eqnarray}
The process of removal of the divergences of this nonrenormalizable 
field theory
\cite{Bettinelli:2007tq},
requires a much wider set of external sources 
($K_0,V_{a\mu},z_0,z_a $)\, than in eq. (\ref{app1.5})
\begin{eqnarray}&&
\int d^Dx \, (A_{a\mu}^* ~ {\mathfrak s} A_a^\mu 
+ \phi_a^* ~{\mathfrak s} \phi_a+ \phi_0^* ~{\mathfrak s} \phi_0 
 + c_a^* ~{\mathfrak s} c_a 
+V_{a\mu}{\mathfrak s} D^\mu[A]_{ab}\bar c_b
\nonumber\\&&
+ z_0\chi_0+z_a\chi_a
+K_0\phi_0 ) \, ,
\label{app2.2}
\end{eqnarray}
where
\begin{eqnarray}&&
\bar c\equiv  \bar c_a \frac{\tau_a}{2}
\nonumber\\&&
\chi_0 + i \tau_a \chi_a \equiv  2i\, {\mathfrak s}\,
\bar c \, \Omega = 2i( b~\Omega -\, \bar c ~{\mathfrak s}~\Omega).
\label{app2.3}
\end{eqnarray}
By construction, the fields
\begin{eqnarray}&&
\chi_0 = - b_a\phi_a + \frac{1}{2}\bar c_a c_a \phi_0+ \frac{1}{2}\epsilon_{abc}\bar c_b c_a \phi_c
\nonumber\\&&
\chi_c = \phi_0 b_c - \epsilon_{abc}b_a \phi_b 
+ \frac{1}{2}\epsilon_{abc}\bar c_b c_a \phi_0
+ \frac{1}{2}\bigl(
\bar c_c c_a\phi_a + \bar c_a c_a \phi_c - \bar c_a \phi_a c_c
\bigr) .
\label{app2.4}
\end{eqnarray}
transform like $\phi_0, \phi_a$. 
The source $K_0$ is needed in order
to perform the insertion of the composite operator $\phi_0$ \cite{Ferrari:2005ii}, $\phi_0^*$ 
and $V_{a\mu}$ are the external sources for the BRST transform of $\phi_0$ 
and for some operator entering in the gauge fixing \cite{Bettinelli:2007tq} and finally
$z_0,z_a$ are necessary in order to deal with the 't Hooft gauge 
\cite{Ferrari:2010ge}.
\par
In a standard way one gets the STI 
\begin{eqnarray}
&&
\int d^Dx \, \Big (
 \Gamma_{ A^*_{a\mu}}  \Gamma_{ A_a^\mu}
+
 \Gamma_{ \phi_a^*}  \Gamma_{ \phi_a}
-K_0
 \Gamma_{ \phi_0^*}  
+ 
 \Gamma_{ c_a^*} \Gamma_{ c_a}
+ b_a  \Gamma_{ \bar c_a} 
\Big ) = 0 \, ,
\label{app2.5}
\end{eqnarray}
and the gauge fixing equation
\begin{eqnarray}
\Gamma_{b_a}
= \Lambda_g\Big[
 \frac{b_a}{\xi }+2\frac{ M^2}{\xi }\phi_a +
 \partial^\nu A_a^\mu \Big] - z_0 \phi_a +z_c (\Gamma_{K_0} -\epsilon_{abc} \phi_b).
\label{app2.6}
\end{eqnarray}
\par
For the connected amplitudes we have
\begin{eqnarray}
\int d^Dx \, \Big (
- J_{a\mu}  W_{ A^*_{a\mu}}
-
 K_a W_{ \phi_a^*} 
- K_0   W_{ \phi_0^*} 
+ 
 \bar\eta_a W_{ c_a^*}
-\eta_a W_{  b_a} 
\Big ) = 0 
\label{app2.7}
\end{eqnarray}
and the gauge-fixing equation
\begin{eqnarray}
-J_{b_a}
=\Lambda_g\Big[
 \frac{1}{\xi }W_{b_a}+2\frac{ M^2}{\xi }W_{\phi_a} +
 \partial^\nu W_{ A_a^\mu} \Big] - z_0 W_{\phi_a} 
+z_c (W_{K_0} -\epsilon_{abc} W_{\phi_b}).
\label{app2.8}
\end{eqnarray}
The presence of the sources $z_a$ upsets the anti-ghost equation, in fact 
$\Gamma_{\bar c_a}$ contains the insertion of composite operators that
are not associated to the listed external sources.  
The use of the sources $K_0,V_{a\mu},z_0,z_a $ is necessary
for the subtraction of the infinities \cite{Ferrari:2010ge}. In this work we evaluate
only the two-point function of $b, A_\mu, \phi$, then we can put to zero
all the external source $K_0,V_{a\mu},z_0,z_a $ and consequently 
the usual anti-ghost equation (see eqs. (\ref{app1.13}) and (\ref{app1.14})) is at our disposal.
\par
In this subsection we have shown that in the massive Yang-Mills
theory the two-point function $W$ and $\Gamma$  in the unphysical sector 
obey the same equations with the Higgs mechanism
as well as with the St\"uckelberg term, apart from an
unessential  rescaling of the field $\phi\to 2M \phi$.
Also the results of Appendix \ref{app:conv} apply, with the same rescaling.
%
\section{Properties of the Two-point Functions (Higgs)}
\label{sec:tpf}
The results of this section apply to both
Higgs and St\"uckelberg scenario, since the
STI (eqs. (\ref{app1.6}), (\ref{app1.8}), (\ref{app2.5})   
and (\ref{app2.7}) ) and the gauge fixing equations 
(eqs. (\ref{app1.11}), (\ref{app1.12}), (\ref{app2.6})   
and (\ref{app2.8}) ) are the
same after rescaling the $\phi$ field.
\par
Now we derive some consequences of the above equations.
\par\noindent
From eq. (\ref{app1.11}) we get
\begin{eqnarray}
\Gamma_{bb} = \Lambda_g\frac{1}{\xi}, \qquad
\Gamma_{b\phi} = \Lambda_g\frac{M}{\xi}, \qquad
\Gamma_{A^\mu b} =   i \Lambda_g p_\mu
\label{app1.14.1}
\end{eqnarray}
From the STI in eq. (\ref{app1.8})
\begin{eqnarray}&&
W_{b_a b_b} = 0
\nonumber\\&&
W_{A_{a\mu}^* \bar c_b } =W_{A_{a\mu} b_b}
\nonumber\\&&
W_{\phi_a^* \bar c_b } =W_{\phi_a b_b}
\label{app1.15}
\end{eqnarray}
and from eq. (\ref{app1.12})
\begin{eqnarray}
\Lambda_g\Big[
\frac{M}{\xi}W_{\phi_a^* \bar c_b } -ip^\mu W_{A_{a\mu}^* \bar c_b }\Big] =-\delta_{ab}.
\label{app1.15.1}
\end{eqnarray}
From the STI in eq. (\ref{app1.13})
\begin{eqnarray}
 \Gamma_{c_a A^*_{{a'}\mu}}  \Gamma_{ b_b A_{a'}^\mu}+\Gamma_{c_a \phi^*_{{a'}}}  \Gamma_{ b_b \phi_{a'}}
+\Gamma_{c_a \bar c_b}  = 0 
\label{app1.16}
\end{eqnarray}
i.e.
\begin{eqnarray}
\Lambda_g\Big[
 - i p_\mu\Gamma_{c_a A^*_{b\mu}}  + \frac{M}{\xi} \Gamma_{c_a \phi^*_{b}} \Big]
=-\Gamma_{c_a \bar c_b}
\, .
\label{app1.17}
\end{eqnarray}
Moreover from the STI in eq. (\ref{app1.6}) we get (we drop unnecessary indexes)
\begin{eqnarray}&&
\Gamma_{c A^*_{\mu}} \Gamma_{ A^\mu \phi}+ \Gamma_{c \phi^*} \Gamma_{\phi \phi}=0
\label{app1.18}
\\&&
\Gamma_{c A^*_{\mu}} \Gamma_{ A^\mu A^\nu}  + \Gamma_{c \phi^*} \Gamma_{\phi A^\nu} = 0 
\, .
\label{app1.19}
\end{eqnarray}
Eqs.  (\ref{app1.18}) and  (\ref{app1.19})  are compatible
if the Jacobian  is zero
\begin{equation}
{
(p^\nu\Gamma_{A^\nu\phi})^2 +p^2\Gamma_L\Gamma_{\phi\phi}=0.
}
\label{tpf.9.2}
\end{equation}
 Now we solve the
linear system given by eqs.  (\ref{app1.17}) and  (\ref{app1.18})
\begin{eqnarray}&&
\Lambda_g p_\mu \Gamma_{c A^*_{\mu}}=- \frac{\Gamma_{c \bar c}\Gamma_{\phi \phi}}{i\Gamma_{\phi \phi}-
 \frac{M}{\xi p^2}p^\nu \Gamma_{ A^\nu \phi}}
= i p^2 \frac{\Gamma_{c \bar c}}{  p^2+i
 \frac{M}{\xi\Gamma_{\phi \phi}}p^\nu \Gamma_{ A^\nu \phi}}
\nonumber\\&& 
\Lambda_g \Gamma_{c\phi^*} = - \frac{p^\nu \Gamma_{ A^\nu \phi}}{ p^2\Gamma_{\phi \phi}}p_\mu \Gamma_{c A^*_{\mu}}
=  - i\frac{p^\nu \Gamma_{ A^\nu \phi}}{ \Gamma_{\phi \phi}}   \frac{\Gamma_{c \bar c}}{  p^2+i
 \frac{M}{\xi\Gamma_{\phi \phi}}p^\nu \Gamma_{ A^\nu \phi}}.
\label{app1.20}
\end{eqnarray}
Thus eq. (\ref{app1.15}) gives
\begin{eqnarray}&&
 W_{A^\mu b}=-\frac{i}{\Lambda_g}\frac{p^\mu}{
p^2-i  \frac{M}{\xi}\frac{p^\nu\Gamma_{\phi A^\nu}}{\Gamma_{\phi\phi}}
}
\nonumber\\&&
 W_{\phi b}
=\frac{i}{\Lambda_g}\frac{p^\nu\Gamma_{\phi A^\nu}}{\Gamma_{\phi\phi}}
\frac{1}{p^2-\frac{M}{\xi}\frac{ip^\nu\Gamma_{\phi A^\nu}}{\Gamma_{\phi\phi}}
}.
\label{basic.8}
\end{eqnarray}
The two-point functions have the pole in the same position, i.e. the solution 
of
\begin{eqnarray}
p^2=\frac{M}{\xi}\frac{ip^\nu\Gamma_{\phi A^\nu}}{\Gamma_{\phi\phi}}.
\label{basic.99p}
\end{eqnarray}
%
\subsection{Two-point Functions}
From the STI eq. (\ref{app1.8}) one gets ($k>1$)
\begin{eqnarray}
W_{b_1\cdots b_k ***}=0,
\label{basic.1}
\end{eqnarray}
where $\scriptstyle{ ***}$ indicates any reduction formula operator
(on-shell) for \underline{physical} modes.
\par\noindent
From eqs.  (\ref{app1.12}) and (\ref{basic.1})
we get 
\begin{eqnarray}&&
\frac{M}{\xi} W_{\phi b}  - i p_\mu W_{A^\mu b} = -\frac{1}{\Lambda_g}
\label{basic.6.1}
\\&&
\frac{1}{\xi} W_{b\phi }  +\frac{M}{\xi} W_{\phi\phi }
 - i p_\mu W_{A^\mu \phi} = 0
\label{basic.6.2}
\\&&
\frac{1}{\xi} W_{b A^\nu } +\frac{M}{\xi} W_{\phi A^\nu }
 - i p_\mu W_{A^\mu A^\nu} = 0.
\label{basic.6.3}
\end{eqnarray}
%
Now we use 
\begin{eqnarray}
\Gamma W = - I\!\!I.
\label{basic.7}
\end{eqnarray}
and eqs. (\ref{basic.1})-(\ref{basic.6.3}) in order
to derive the two-point function $W$ in terms
of the two-point function $\Gamma$.
We explicit write some elements of the
matrix in eq. (\ref{basic.7})
\begin{eqnarray}&&
\bigl(\Gamma W\bigr)_{\phi b}=0\Longrightarrow \Gamma_{\phi\phi}W_{\phi b}
+\Gamma_{\phi A^\mu}W_{A^\mu b}=0
\label{tpf.8.1}
\\&&
\bigl(\Gamma W\bigr)_{A^\nu b}=0\Longrightarrow \Gamma_{A^\nu\phi}W_{\phi b}
+\Gamma_{A^\nu A^\mu}W_{A^\mu b}=0
\label{tpf.8.2} \\&&
\bigl(\Gamma W\bigr)_{\phi \phi}=-1\Longrightarrow \Gamma_{\phi b}W_{b \phi}
+\Gamma_{\phi\phi}W_{\phi \phi} +\Gamma_{\phi A^\mu}W_{A^\mu \phi}=-1
\label{tpf.8.3} \\&&
\bigl(\Gamma W\bigr)_{\phi A^\nu }=0\Longrightarrow \Gamma_{\phi b}W_{ bA^\nu}+
\Gamma_{\phi\phi}W_{\phi A^\nu}+\Gamma_{\phi A^\mu}W_{A^\mu A^\nu}=0
\label{tpf.8.4} \\&&
\bigl(\Gamma W\bigr)_{ A^\nu \phi }=0\Longrightarrow \Gamma_{A^\nu b}W_{ b\phi}+
\Gamma_{ A^\nu\phi}W_{\phi\phi}+\Gamma_{ A^\nu A^\mu}W_{A^\mu\phi}=0
\label{tpf.8.5} \\&&
\bigl(\Gamma W\bigr)_{ A^\nu A^\mu }=-g_{\mu\nu}
\nonumber\\&&
\Longrightarrow \Gamma_{A^\nu b}W_{ bA^\mu}+
\Gamma_{ A^\nu\phi}W_{\phi A^\mu}+\Gamma_{ A^\nu A^\sigma}W_{A^\sigma A^\mu}
=-g_{\mu\nu}
\label{tpf.8.6}
\end{eqnarray}
\par\noindent
Now we solve the linear system given by the eqs.(\ref{basic.6.2}) and  (\ref{tpf.8.3}). 
The jacobian of the homogeneous system is
\begin{eqnarray}
J_\Delta = p^\nu\Gamma_{\phi A^\nu}\frac{M}{\xi p^2} + i\Gamma_{\phi\phi}
=i\frac{\Gamma_{\phi\phi}}{ p^2}\Bigl( p^2-ip^\nu\Gamma_{\phi A^\nu}\frac{M}{\xi 
\Gamma_{\phi\phi}}\Bigr)
\label{tpf.23.-1}
\end{eqnarray}
The straightforward solutions are
\begin{eqnarray}
  W_{A^\mu \phi}
=i \frac{1}{\xi\Gamma_{\phi\phi}}\biggl(\frac{i}{\Lambda_g}p^\nu\Gamma_{ A^\nu\phi}
+M p^2
\biggr)
\frac{p^\mu
}{\biggl(
p^2 -  \frac{M}{\xi} 
\frac{ip^\nu\Gamma_{ \phi A^\nu}}{\Gamma_{\phi\phi}}\biggr)^2
}
\label{tpf.23}
\end{eqnarray}
and
\begin{eqnarray}
W_{\phi \phi}
 =-
\frac{p^2}{ \Gamma_{\phi \phi}}
\biggl(p^2-
\frac{1}{\Lambda_g \xi}\Gamma_L 
\biggr)
\frac{1}{\biggl(
p^2 -  \frac{M}{\xi}
\frac{ip^\nu\Gamma_{\phi A^\nu}}{\Gamma_{\phi\phi}}\biggr)^2
}.
\label{tpf.22}
\end{eqnarray}
A similar calculation for the linear system
given by the eqs. (\ref{basic.8}), (\ref{basic.6.3}) and  (\ref{tpf.23}) yields
\begin{eqnarray}
W_L 
=\frac{p^2}{\xi\Gamma_{\phi\phi}}
\frac{\frac{1}{\Lambda_g}\Gamma_{\phi\phi} -\frac{M^2}{\xi}}{\biggl(
p^2 -  \frac{M}{\xi} \frac{ip^\nu\Gamma_{\phi A^\nu}}
{\Gamma_{\phi\phi}}\biggr)^2
}.
\label{tpf.24}
\end{eqnarray}
The presence of double poles is the source of some technical problems in dealing
with the proof of Physical Unitarity.
\par
For comparison, we evaluate the relevant quantities for the free fields
\begin{eqnarray}&&
\Gamma_{bb}= \frac{\Lambda_g}{\xi}, \quad
\Gamma_{\phi b }=\Lambda_g \frac{M}{\xi}, \quad
\Gamma_{ A^\nu b}=i\Lambda_g p_\nu, \quad
\nonumber\\&&
\Gamma_{ A^\nu\phi}=i\Lambda_g Mp_\nu, \quad
\Gamma_{\phi\phi}=\Lambda_g p^2, \quad
\Gamma_L=\Lambda_g M^2.
\label{tpf.25}
\end{eqnarray}
Then
\begin{eqnarray}&&
 W_{A^\mu\phi }=0
\label{tpf.26.1}
\end{eqnarray}
and
\begin{eqnarray}&&
 W_{A^\mu b}=-i\frac{p^\mu}{\Lambda_g(
p^2-  \frac{M^2}{\xi})
},\qquad
 W_{\phi b}=\frac{M}{\Lambda_g(
p^2 -  \frac{M^2}{\xi})
}
\nonumber\\&&
 W_L=\frac{1}{\Lambda_g\xi}\frac{1}{
p^2 -  \frac{M^2}{\xi}
},\qquad
W_{\phi \phi}
 =- \frac{1}{\Lambda_g(
p^2 -  \frac{M^2}{\xi})
}.
\label{tpf.27.1}
\end{eqnarray}
Only simple poles appear in the free field approximation.
\section{Zero Eigenvalue Mode}
\label{app:conv}
We construct the mode corresponding to the zero eigenvalue of the matrix ${\cal G}$ in
(\ref{ptpf.26}) when on-shell.

 First consider
\begin{eqnarray}
\Upsilon _a\equiv \frac{1}{\xi}b_a + \frac{M}{\xi}\phi_a + \partial_\mu A_a^\mu.
\label{tpf.41}
\end{eqnarray}
From eq. (\ref{app1.12}) we see that the field decouples
from every other 
\begin{eqnarray}
W_{\Upsilon b}=-1, \qquad W_{\Upsilon \phi}=0, \qquad W_{\Upsilon A_\mu}=0.
\label{tpf.42}
\end{eqnarray}
A next interesting local field is (used in Ref. \cite{Becchi:1974md})
\begin{eqnarray}
X_1\equiv \frac{1}{\xi} \phi
 + \frac{1}{M}\partial_\mu A^\mu. 
\label{tpf.48}
\end{eqnarray}
From eqs. (\ref{basic.6.2}), (\ref{basic.6.3}) and (\ref{basic.8}) we have
\begin{eqnarray}&&
W_{X_1\phi} = - \frac{1}{\xi M}W_{b\phi}
=-\frac{i}{\Lambda_g }\frac{p^\nu\Gamma_{\phi A^\nu}}{\xi M\Gamma_{\phi\phi}}
\frac{1}{p^2-\frac{M}{\xi}\frac{ip^\nu\Gamma_{\phi A^\nu}}{\Gamma_{\phi\phi}}
}
\nonumber\\&&
W_{X_1 A^\nu} = - \frac{1}{\xi M}W_{b A^\nu}
=-\frac{i}{\Lambda_g \xi  M}\frac{p^\nu}{
p^2-i  \frac{M}{\xi}\frac{p^\nu\Gamma_{\phi A^\nu}}{\Gamma_{\phi\phi}}
}.
\label{tpf.49}
\end{eqnarray}
From eq. (\ref{tpf.49}) we see that $X_1$ is the mode corresponding to the zero
eigenvalue of $\cal G$ when taken on-shell.
Finally one has
\begin{eqnarray}&&
W_{X_1X_1}= \frac{1}{\xi}W_{X_1\phi}+i\frac{1}{M}p^\nu W_{X_1 A^\nu}
\nonumber\\&&
=-\frac{1}{\Lambda_g\xi  M^2}\biggl(i\frac{M}{\xi}\frac{p^\nu\Gamma_{\phi A^\nu}}{\Gamma_{\phi\phi}}
-p^2
\biggr)
\frac{1}{p^2-\frac{M}{\xi}\frac{ip^\nu\Gamma_{\phi A^\nu}}{\Gamma_{\phi\phi}}
}
= \frac{1}{\Lambda_g\xi M^2}.
\nonumber\\&&
\label{tpf.50}
\end{eqnarray}
%

\end{document}